\documentclass[12pt]{article}
\usepackage{xcolor}
\usepackage{cite}
  \usepackage{graphicx}
  \usepackage{epsfig}
  \usepackage{amssymb}
  \usepackage{subfigure}
  \usepackage{physics}
  \usepackage{amsmath} 

\evensidemargin - .5truecm
\allowdisplaybreaks[1]

\begin{document}

\newcommand{\lsim}{\lesssim}
\newcommand{\gsim}{\gtrsim}

\newcommand{\CC}{{\mathbb C}}
\newcommand{\RR}{{\mathbb R}}
\newcommand{\ZZ}{{\mathbb Z}}
\newcommand{\QQ}{{\mathbb Q}}
\newcommand{\NN}{{\mathbb N}}
\newcommand{\beq}{\begin{equation}}
\newcommand{\eeq}{\end{equation}}
\newcommand{\beal}{\begin{align}}
\newcommand{\eeal}{\end{align}}
\newcommand{\nn}{\nonumber}
\newcommand{\bea}{\begin{eqnarray}}
\newcommand{\eea}{\end{eqnarray}}
\newcommand{\ba}{\begin{array}}
\newcommand{\ea}{\end{array}}
\newcommand{\bfig}{\begin{figure}}
\newcommand{\efig}{\end{figure}}
\newcommand{\bc}{\begin{center}}
\newcommand{\ec}{\end{center}}

\newenvironment{appendletterA}
{
  \typeout{ Starting Appendix \thesection }
  \setcounter{section}{0}
  \setcounter{equation}{0}
  \renewcommand{\theequation}{A\arabic{equation}}
 }{
  \typeout{Appendix done}
 }
\newenvironment{appendletterB}
 {
  \typeout{ Starting Appendix \thesection }
  \setcounter{equation}{0}
  \renewcommand{\theequation}{B\arabic{equation}}
 }{
  \typeout{Appendix done}
 }

\begin{titlepage}
\nopagebreak

\renewcommand{\thefootnote}{\fnsymbol{footnote}}
\vskip 2cm
%\begin{center}
%\boldmath

\vspace*{1cm}
\begin{center}
{\Large \bf 
  Thrust distribution in electron-positron annihilation
\\[0.1cm]
 at full N$\bf ^3$LL+NNLO (and beyond) in QCD \\[0.1cm]
}
\end{center}

%{\Large\bf 

\par \vspace{1.5mm}
\begin{center}
  {\bf Ugo Giuseppe Aglietti${}^{(a)}$},  {\bf Giancarlo Ferrera${}^{(b,c)}$}, {\bf  Wan-Li Ju${}^{(c,d)}$} and {\bf  Jiahao Miao${}^{(b)}$\!\!\!\!}
\vspace{5mm}

${}^{(a)}$
Dipartimento di Fisica, Universit\`a di Roma ``La Sapienza'' and\\ INFN, Sezione di Roma,
I-00185 Rome, Italy\\\vspace{1mm}

${}^{(b)}$ 
Dipartimento di Fisica, Universit\`a di Milano,
I-20133 Milan, Italy\\\vspace{1mm}

${}^{(c)}$ 
INFN, Sezione di Milano,
I-20133 Milan, Italy\\\vspace{1mm}

${}^{(d)}$ 
Department of Physics, University of Alberta, Edmonton AB T6G 2J1, Canada

\end{center}

\vspace{.5cm}

%\pacs{13.87.Ce,12.38Bx}

\par %\vspace{2mm}
\begin{center} {\large \bf Abstract} \end{center}
\begin{quote}
\pretolerance 10000
%\begin{abstract}

We consider the  thrust ($T$) distribution 
in  electron-positron ($e^+e^-$) annihilation into hadrons and we perform the all-order resummation of the large logarithms of $1-T$
up to next-to-next-to-next-to-next-to-leading logarithmic (N$^4$LL) accuracy in QCD.
We consistently combine resummation with the known fixed-order results up to next-to-next-to-leading order (NNLO). 
All perturbative terms up to $\mathcal{O}(\alpha_S^3)$  are included in our calculation, which, thanks to a unitarity constraint, exactly reproduces, after integration over $T$,
the next-to-next-to-next-to-leading order (N$^3$LO) result for the total cross section of $e^+e^-$ into hadrons.
%We regularize the Landau singularity of the QCD coupling within the so-called Minimal Prescription. 
We perform resummation in the Laplace-conjugated space and compare our results with those obtained with the resummation formalism in the physical ($T$) space.
We find that the differences in the spectra obtained with the two different formalisms are sizable.
%We find that the numerical differences of the two approaches are sizable.
%We exhibit and discuss the reduction of the perturbative scale dependence of  distributions at higher orders, as a means to estimate the corresponding residual perturbative uncertainty.
Non-perturbative corrections are included using an analytic hadronization model, depending on two free parameters.
Finally, we present a comparison of our spectra
with experimental data at the $Z$-boson mass ($m_Z$) energy, which enables us to extract the value of the QCD coupling $\alpha_S(m_Z^2)=0.1181 \pm 0.0018$ fully consistent
with the world average.
\vskip .4cm

%\end{abstract}
\vfill\end{quote}
%\vspace*{1.cm}
%\begin{center}
%\end{center}
%\vspace*{1cm}
%\vspace*{\fill} 

\begin{flushleft}
February 2025
\vspace*{-1cm}
\end{flushleft}
\end{titlepage}

\renewcommand{\thefootnote}{\fnsymbol{footnote}}

\newpage

%%%%%%%%%%%%%%%%%%%%%%

%\section{Introduction}

%%%INTRODUCTION

One of the key parameters of the Standard Model is the QCD coupling constant $\alpha_S$, which governs the
strength of the strong interactions. While the nonabelian gauge structure of QCD predicts the (logarithmic) energy scaling of the coupling via the
renormalization group equation, the value of the coupling at an initial (or reference) scale has to be determined by comparing theory predictions against experimental data.
Among the fundamental couplings, $\alpha_S$ is by far the least well-known. Increasing its accuracy,
%of $\alpha_S$ at various energy scale
being {\itshape per se} interesting to test the over-all consistency of QCD,
%Moreover the value of  and its knowledge
is crucial for improving the theoretical predictions for hard scattering processes at, present and future, high-energy
 colliders, such as the Large Hadron Collider (LHC).

A classical method to obtain a precise determination of $\alpha_S$
involves the analysis of  shape variables  in high-energy
electron-positron annihilation into hadrons.  Shape variables are functions of the space momenta of all the particles in the final state, which characterize the topology
of an event. Shape variables are strongly sensitive to QCD radiation and thus to the precise value of the strong coupling, offering then a method
for an accurate extraction of $\alpha_S$.
Among the first and most widely used shape variables is the thrust $T$\,\cite{Farhi:1977sg}   defined as
\begin{equation}
\label{thrust}
T %\equiv{\rm max}_{\bf n} T_{\bf n}
\equiv 1-\tau = {\rm max}_{\bf n} \frac{\sum_i|{\bf p}_i \cdot {\bf n}|}{\sum_i |{\bf p}_i|}\,,
\end{equation}
where the sum is over all final state particles $i$ with three-momentum ${\bf p}_i$ and the maximum is taken with respect to the direction of the unit three-vector ${\bf n}$.
The thrust variable thus maximizes the sum of the moduli of the projected three-momenta along the vector ${\bf n}$, and the value of ${\bf n}$ that realizes this maximum
is called the thrust axis ${\bf n}_T$.
As is clear from its definition in Eq.\,(\ref{thrust}), the thrust is an infrared (soft and collinear) safe quantity, that is, it
is insensitive to the emission of zero-momentum (soft) particles and to the splitting of one particle into two collinear ones. Therefore the thrust distribution
can be safely calculated in a perturbative QCD
expansion in powers of $\alpha_S$. % neglecting non-perturbative (NP) power suppressed contributions.
It can easily be shown that the allowed values of the thrust lie in the range $1/2 \leq T \leq 1$: it approaches $T=1$ ($\tau=0$) in the two-jet limit (pencil-like event)
and $T=1/2$ ($\tau=1/2$) in the opposite limit of a completely isotropic event.

Away from the two-jet region, the perturbative
QCD series for the thrust distribution is well behaved, and
calculations based on the truncation at a fixed order in $\alpha_S$
are reliable. Since hadron production away from the back-to-back region
has to be accompanied by the radiation of at least one hard recoiling parton,
the leading-order (LO) term for this observable is $\mathcal{O}(\alpha_S)$. 
The fixed-order expansion of the thrust distribution for $T \neq 1$ is known up ${\mathcal O}(\alpha_S^3)$. It has been obtained
by numerical Monte Carlo integration of the fully differential cross section for three-jet production in electron-positron annihilation
at next-to-next-to-leading (NNLO) order in QCD\,\cite{Gehrmann-DeRidder:2007nzq,Weinzierl:2008iv,Gehrmann-DeRidder:2009fgd,Weinzierl:2009ms,Gehrmann-DeRidder:2014hxk,DelDuca:2016csb,DelDuca:2016ily}.

However at large values of $T$ ($\tau=1-T \ll 1$), the fixed-order QCD perturbative expansion does not provide
a good approximation, because its coefficients contain large logarithmic corrections of infrared (soft and collinear) origin, of the type
$\ln\tau$.
The resummation of such
logarithms to all orders in $\alpha_S$  has been formulated in 
the seminal papers \cite{Catani:1991kz,Catani:1992ua}. More recent studies can be found in
Refs.\,\cite{Schwartz:2007ib, Becher:2008cf, Dissertori:2009ik, Hornig:2009vb, Abbate:2010xh, Monni:2011gb, Abbate:2012jh,Almeida:2014uva,Banfi:2014sua,Baron:2018nfz,Wang:2019ljl,Marzani:2019evv,Benitez:2024nav}.

Resummed and fixed-order calculations must be consistently combined
with each other at intermediate values of $T$,
in order to obtain accurate QCD predictions over a wide kinematical region.

In this Letter, we present a resummed calculation of the infrared (or Sudakov) logarithms
occurring in the thrust distribution in the back-to-back region,  
up to next-to-next-to-next-to-next-to-leading logarithmic (N$^4$LL)
accuracy in full QCD. We also match the resummed calculation with the corresponding  NNLO results in
Ref.\,\cite{Weinzierl:2009ms}.
Therefore all perturbative terms up to the next-to-next-to-next-to-leading order
(N$^3$LO), i.e.\ up to $\mathcal{O}(\alpha_S^3)$, are consistently included in our calculation.

We implement a unitarity constraint in the resummation formalism %following Ref.\,\cite{},
in such a way that our calculation exactly reproduces, 
after integration over $T$, the corresponding fixed-order result for the total cross section of electron-positron annihilation into hadrons
up to N$^3$LO\,\cite{Gorishnii:1990vf,Surguladze:1990tg}.

We have then  included non-perturbative (NP) effects in the thrust distribution\,\cite{Webber:1994cp,Dokshitzer:1995zt,Nason:1995np,Dokshitzer:1997ew,Gardi:1999dq,Gardi:2000yh,Gardi:2001ny,Lee:2006nr}, by using an analytic hadronization model depending on two free parameters,
which enables us to perform an  extraction of the value of the QCD coupling,
by comparing our results with LEP and SLD data at the $Z$-boson mass ($m_Z$) energy.

The result of our fit, $\alpha_S(m_Z^2)=0.1181 \pm 0.0018$,  is fully consistent
with the world average\,\cite{ParticleDataGroup:2024cfk}.
Furthermore, we observe that our result is substantially higher than other determinations of $\alpha_S$   based on thrust-space resummation%
\,\cite{Abbate:2010xh,Gehrmann:2012sc,Benitez:2024nav}\,\footnote{See also Refs.\,\cite{Nason:2023asn,Nason:2025qbx}
and the discussion in the Sec.\ 9.4.4 of Ref.\,\cite{ParticleDataGroup:2024cfk}.}.
We will explicitly show that, performing the resummation of Sudakov logarithms in Laplace-conjugated space
(ensuring the factorization of kinematical momentum conservation constraint), is crucial to obtain a result which is consistent with the world
average.

We start considering the differential thrust distribution $d\sigma/d\tau$ and we define the {\itshape cumulative} distribution
\beq
\label{cumu}
R_T(\tau) \equiv \frac1{\sigma_{\mathrm{tot}}} \int_0^\tau d\tau' \frac{d\sigma}{d\tau'}\,,
\eeq
where $\sigma_{\mathrm{tot}}$ is the total cross section of $e^+e^-$ annihilation into hadrons.
%%%%%% FIXED ORDER

The upper limit of $\tau$, $\tau_{\max}$, (or, equivalently, the lower limit on $T$, $T_{\min}=1-\tau_{\max}$)
depends on the number ($n$) of final-state partons and approaches from below the value $\tau_{\max}=1/2$, in the (formal) limit of infinite parton emissions ($n\to \infty$).
The knowledge of $\tau_{\max}$ is important in order to correctly normalize the cross section.
In the massless approximation, for three particles ($n=3$), $\tau_{\max}=1/3$, corresponding to a symmetric trigonal-planar configuration.
For four particles ($n=4$), $\tau_{\max}=1-1/\sqrt{3}=0.4226497\cdots$, which corresponds to final-state three-momenta forming
a regular tetrahedron. For more than four particles, as far as we know,
there are no known values of $\tau_{\max}$ in the literature. For five particles ($n=5$), $\tau_{\max}$ was previously estimated to be
$\tau_{\max}= 0.4275$\,\cite{Monni:2011gb}. The estimate was obtained
by considering at which bin the cross section of the NNLO Monte Carlo (MC) calculation\cite{Gehrmann-DeRidder:2014hxk} vanishes.

In general, finding the upper limit of $\tau$ for a generic number $n$ of final state particles is a non-trivial kinematical problem, as it involves a double optimization.
In this paper, to find the maximum value of $\tau$  for a given $n$, we have used stochastic optimization algorithms,
such as the Genetic Algorithm  and the Particle Swarm Optimization.
The results for the $3\leq n \leq 14$ are shown in Tab.\,(1). The numbers in the table should be interpreted as lower limits to the maximum
value of $\tau$, since a stochastic algorithm is not guaranteed to find the absolute minimum. However, we observe that the results obtained
with these algorithms for $n=3$ and $n=4$ agree (within the numerical precision of ${\mathcal O}(10^{-4})$) with the exact analytical values.
\begin{table}[ht]
  \begin{center}
    \begin{tabular}{||l|l|l|l|l|l|l||}
  \hline\hline
  $n$   & 3         & 4         & 5         & 6         & 7         & 8         \\\hline
  %$\tau_{\max}$ & 0.33333       & 0.42265     & 0.4539536 & 0.4616661 & 0.4716270 & 0.4737373 \\\hline\hline
$\tau_{\max}$ & 0.3333       & 0.4226     & 0.4539 & 0.4629 & 0.4716 & 0.4753 \\\hline\hline  
$n$   & 9         & 10        & 11        & 12        & 13        & 14        \\\hline
%  $\tau_{\max}$ & 0.4782286 & 0.4802864 & 0.4816904 & 0.4840608 & 0.4836307 & 0.4842613\\
$\tau_{\max}$ & 0.4790 & 0.4811 & 0.4834 & 0.4842 & 0.4845 & 0.4857\\  
  \hline\hline
\end{tabular}
\caption{
{\em
Maximum kinematically allowed value of $\tau$ as a function of the number of particles $n$ in
the final state. }}
\end{center}
\label{tab1}
\end{table}

The cumulant cross section $R_T(\tau)$ in Eq.\,(\ref{cumu}) can be factorized as follows\,\cite{Catani:1991kz,Catani:1992ua}:
\begin{equation}\label{eq:cumulant_expansion}
    R_T(\tau) =  C\left(\alpha_S(Q^2)\right)\Sigma\left(\tau,\alpha_s(Q^2)\right) + D\left(\tau,\alpha_s(Q^2)\right).
\end{equation}
$C(\alpha_S)$ is a hard-virtual factor with a standard (fixed-order) perturbative expansion:
\begin{equation}\label{Chard}
   C(\alpha_S)= 1+ \sum_{n=1}^{\infty} \left(\frac{\alpha_S}{\pi}\right)^n C_n\,.
\end{equation}
$D(\alpha_S)$ is a short-distance {\itshape remainder} function vanishing at small $\tau$,
\begin{equation}\label{Drem}
   D(\tau,\alpha_S)= \sum_{n=1}^{\infty} \left(\frac{\alpha_S}{\pi}\right)^n D_n(\tau)\,.
\end{equation}
$\Sigma\left(\tau,\alpha_s\right)$ is a long-distance dominated form factor, which contains all the logarithms of soft and collinear origin enhanced at small $\tau$,
of the type $\alpha_S^n\ln^m\tau$ ($1 \leq m\leq 2n$); it can be written in an exponential form in the
Laplace-conjugated space:
\begin{equation}\label{sigma}
   \Sigma(\tau,\alpha_S)= \frac1{2\pi i}\int_{C}\frac{dN}{N} e^{N\tau} e^{{\mathcal F}(\alpha_S,L)}\,,
\end{equation}
where the contour of integration $C$ runs parallel to the imaginary axis and lies to the
right of all singularities of the integrand. The exponent ${\mathcal F}(\alpha_S L)$ in Eq.\,(\ref{sigma})
  resums to all orders  in $\alpha_S$, classes of logarithms $L=\ln N$, which are large for $N\to \infty$ (and correspond, in physical space, to
$\ln \tau$ terms, enhanced in the two-jet region $\tau \to 0$):
  \begin{equation}\label{ffun}
    {\mathcal F}(\alpha_S,L)=L\,f_1(\lambda)+f_2(\lambda)+\frac{\alpha_S}{\pi}f_3(\lambda)+\sum_{n=4}^{\infty} \left(\frac{\alpha_S}{\pi} \right)^{n-2} f_n(\lambda)\,,
  \end{equation}
  with $\lambda\equiv\beta_0\alpha_S L/\pi$ and $\beta_0$ is the first-order coefficient of the QCD $\beta$ function.  
  At large $N$, $L \gg 1$ and $\alpha_S L$ is assumed to be $\mathcal{O}(1)$,
  implying that  the r.h.s.\ of Eq.\,(\ref{ffun}) has a customary perturbative expansion in powers of $\alpha_S$.
The truncation of such a (function) series at a given order yield the resummation of certain classes (infinite series) of 
logarithmic corrections.  
The leading-logarithmic (LL) approximation is provided by the function 
$f_1(\lambda)$, the next-to-LL (NLL) approximation requires also the inclusion of the function
$f_2(\lambda)$,  the next-to-NLL (NNLL) approximation
requires also the function $f_3(\lambda)$  and so on. We have explicitly evaluated the resummation
functions $f_n(\lambda)$, with $1\leq n\leq 5$ (explicit expressions are given in the Appendix), thus reaching the N$^4$LL accuracy.
We have also evaluated up to ${\mathcal O}(\alpha_S^3)$ both the coefficients of the hard factor $C(\alpha_S)$ and,
by using the numerical NNLO results of Ref.\,\cite{Weinzierl:2009ms},
the remainder function $D(\alpha_S)$.
By evaluating numerically the inverse Laplace transform ($N\mapsto \tau$) in Eq.\,(\ref{sigma}), we have been able to perform the resummation of the large logarithms $\ln N$ in
Laplace space (or $N$ space) up to N$^4$LL accuracy, matching the results up to $O(\alpha_S^3)$, i.e.\ with the N$^3$LO hard virtual corrections at small $\tau$ and
the NNLO hard real (and real-virtual) corrections at large $\tau$.
The expression of the factor given in Eq.\,(\ref{sigma}) % only has a formal meaning, as it
involves an integration over the (formally non-integrable) Landau  singularity of the QCD running,
which manifests itself in singularities of the $f_n(\lambda)$ functions at the points $\lambda = 1/2$ and $\lambda = 1$ (i.e.\ $N\sim N_L= \exp{1/[2\beta_0 \alpha_S(Q^2)]}$
and $N\sim N_L' = \exp{1/[\beta_0 \alpha_S(Q^2)]}$, respectively).
These singularities, which signal the onset of non-perturbative
phenomena at very large values of $N$ (i.e.\ in the region of
very small $\tau$), have been regularized through the so-called
Minimal Prescription of Ref.\,\cite{Catani:1996yz},
in which the contour of integration $C$ lies to the right of all physical
singularities but to the left of the (unphysical) Landau pole.
The results obtained by using this prescription converge asymptotically to the perturbative series
and do not include any power correction. %suppressed contributions. % of the type $\Lambda_{QCD}/Q$.

A commonly used alternative to the numerical computation of the inverse Laplace transform,
is an analytic approximate calculation of the latter.
In this case, the Taylor expansion of the integrand in Eq.\,(\ref{sigma})
around the point\,\cite{Catani:1991kz,Catani:1992ua,Aglietti:2002ew} %its saddle-point
\begin{equation}
\label{Nwrong}  
\ln N=  \ln(1/\tau)\equiv \ell \,,
\end{equation}
gives
\begin{equation}\label{sigma2}
  \Sigma(\tau,\alpha_S)=  \frac1{2\pi i}\int_{C}\frac{dN}{N} e^{N\tau}
 \exp\!\left[\sum_{k=0}^\infty\frac{{\mathcal F}^{(k)}(\alpha_S,\ell)}{k!}\ln^k(\tau N)\right]\,, %= e^{{\mathcal G}(\alpha_S,\ell)}\,,
\end{equation}
where 
\begin{equation}\label{effe}
  {\mathcal F}^{(k)}(\alpha_S,\ell)\equiv\frac{\partial^{k}}{\partial\ell^{k}}
  {\mathcal F}(\alpha_S,\ell)\,,\qquad k=1,2,3,\cdots. %= e^{{\mathcal G}(\alpha_S,\ell)}\,,
\end{equation}
Since it is not possible to evaluate  the series on the r.h.s.\ of  Eq.\,(\ref{sigma2})
exactly, a hierarchy is defined in $\tau-$space. %, similar to the one discussed above in $N$-space.
The N$^n$LL accuracy in $\tau$ space is defined by
keeping in Eq.\,(\ref{sigma2}) the dominant logarithmic terms $\alpha_S^{n-1}(\alpha_S\ell)^k$,
up to a given $n$ and for all $k$.
However, we stress that the resulting resummation formula in $\tau$ space is 
only an approximation of the resummation formula in $N$ space. 
In particular, the resummation in $\tau$ space at a given logarithmic accuracy
{\itshape does not} resum all the $\ln N$ terms
of the corresponding logarithmic accuracy in $N$ space.
The crucial point is that there is not an exact correspondence
between $\ln N$ and $\ln(1/\tau)$ terms. 
The kinematical constraints of momentum conservation  factorize only in $N$ space, not in $\tau$ space, and thus exponentiation
is strictly valid only in $N$ space.

Let us also remark that the above analytic approximation is not a fully consistent saddle-point expansion; that implies that the resulting expansion is not a truly asymptotic one for large $N$.
The true saddle point is obtained by
finding the value of $N$ for which the logarithm of the integrand in Eq.\,(\ref{sigma}) is stationary, that is, by solving the following equation in $N$:
\begin{equation}
\label{eq:Ntrue}
N \tau =
1  -  
f_1\Big(\beta_0 \frac{\alpha_S}{\pi} \ln N\Big)
 - 
\beta_0 \frac{\alpha_S}{\pi} \ln(N)
f_1'\Big(\beta_0 \frac{\alpha_S}{\pi} \ln N\Big)  
-  \beta_0 \frac{\alpha_S}{\pi}
%\hat{{\mathcal F}}'\left[ \beta_0 \frac{\alpha_S}{\pi} \log(N) \right]
\sum_{n=2}^{\infty} \left(\frac{\alpha_S}{\pi} \right)^{n-2} f_{n}'\Big(\beta_0 \frac{\alpha_S}{\pi} \ln N\Big)\,,
\end{equation}
where $f_{n}'(\lambda)\equiv df_{n}(\lambda)/d\lambda$.
The value of $N$ in Eq.\,(\ref{Nwrong})
is instead the solution of the saddle point
equation of the {\it free} theory
($\alpha_S\to 0$),
\beq
N=\frac{1}{\tau}.
\eeq
While the last term in the r.h.s.\ of Eq.\,(\ref{eq:Ntrue}) is a $\mathcal{O}(\alpha_S)$
correction to the saddle point of the free theory,
the same is not true for the second
and the third terms, which are both 
$\mathcal{O}(1)$, since $\alpha_S \ln N$ is $\mathcal{O}(1)$.
We have thus proved that the analytic
formula based on the expansion around the
point in Eq.\,(\ref{Nwrong}) is not a fully consistent
saddle point expansion
and therefore not a correct asymptotic expansion
for large $N$.

As observed in Ref.\,\cite{Catani:1992ua},  while the resummation procedure uniquely determines the structure of the large logarithms in the two-jet region, it leaves an ambiguity on
the non-logarithmic contributions, which is partially solved with the matching against the fixed-order result. This ambiguity can be exploited to impose the following
physical constraints:
\begin{equation}\label{unit}
    R_T(\tau_{\max}) =  1 \,,\qquad \frac{dR_T}{d\tau} (\tau=\tau_{\max})=0\,,
\end{equation}
which are violated by higher-order terms beyond the nominal fixed-order accuracy of the calculation (e.g.\ by $\mathcal{O}(\alpha_S^4)$ terms at N$^3$LL+NNLO).
The first requirement in Eq.\,(\ref{unit}), which follows from perturbative unitarity, is particularly relevant. It can be fulfilled in $\tau$ space
using the following replacement\,\cite{Catani:1992ua,Davison:2009wzs}
\begin{equation}\label{unitl}
\ell\equiv\ln(1/\tau)  \quad\mapsto\quad \tilde\ell\equiv\ln(1/\tau-1/\tau_{\max}+1)\stackrel{\tau\ll 1}{=}\ell +\mathcal{O}(\tau)\,,
\end{equation}
which indeed acts as a perturbative unitarity constraint, since:
\begin{equation}\label{units}
  \Sigma(\tau=\tau_{\max},\alpha_S)|_{\ell\to\tilde\ell}=1\,.
\end{equation}
An analogous constraint can be imposed in $N$ space by means of the following replacement
\begin{equation}\label{unitl2}
L\equiv\ln N \quad\mapsto\quad \tilde L \equiv \ln (N-N_c)\stackrel{N\gg 1}{=}L +\mathcal{O}(1/N)\,,
\end{equation}
with $N_c$ a constant %of $\mathcal{O}(1)$
such that
\begin{equation}\label{units2}
  \Sigma(\tau=\tau_{\max},\alpha_S)|_{L\to\tilde L}=1\,.
\end{equation}
We observe that the replacements in Eqs.\,(\ref{unitl},\ref{unitl2}), besides affecting $\Sigma(\tau,\alpha_S)$, have also an impact on the remainder function $D(\tau,\alpha_S)$, which has to be properly taken into account.  
The second requirement in Eq.\,(\ref{unit})  can be fulfilled by a suitable modification of subleading terms away from the $\tau\to 0$ limit, as discussed in Ref.\,\cite{Catani:1991kz,Davison:2009wzs}.
However, the second requirement is mainly relevant in the large-$\tau$ region ($\tau\lsim \tau_{\max}$), where the thrust distribution
is affected by instabilities from higher-order enhanced corrections of soft and collinear origin near the kinematic fixed-order boundaries\,\cite{Catani:1997xc}.
A proper treatment of these regions requires a resummation of such corrections to all orders, which gives rise to the so called
{\itshape Sudakov shoulder}\,\cite{Catani:1997xc,Aglietti:2001br,Aglietti:2005mb,Bhattacharya:2022dtm}.

In the following part of this Letter, we present Laplace-space ($N$-space) resummed results and we compare them with corresponding results in  $\tau$ space. We will show that the formally subleading
terms neglected in the $\tau$-space resummation, actually give a non negligible contribution and turn out to be essential for a precise extraction of the value of $\alpha_S$.

\begin{figure}[ht]
\begin{center}
\includegraphics[width=0.8\textwidth]{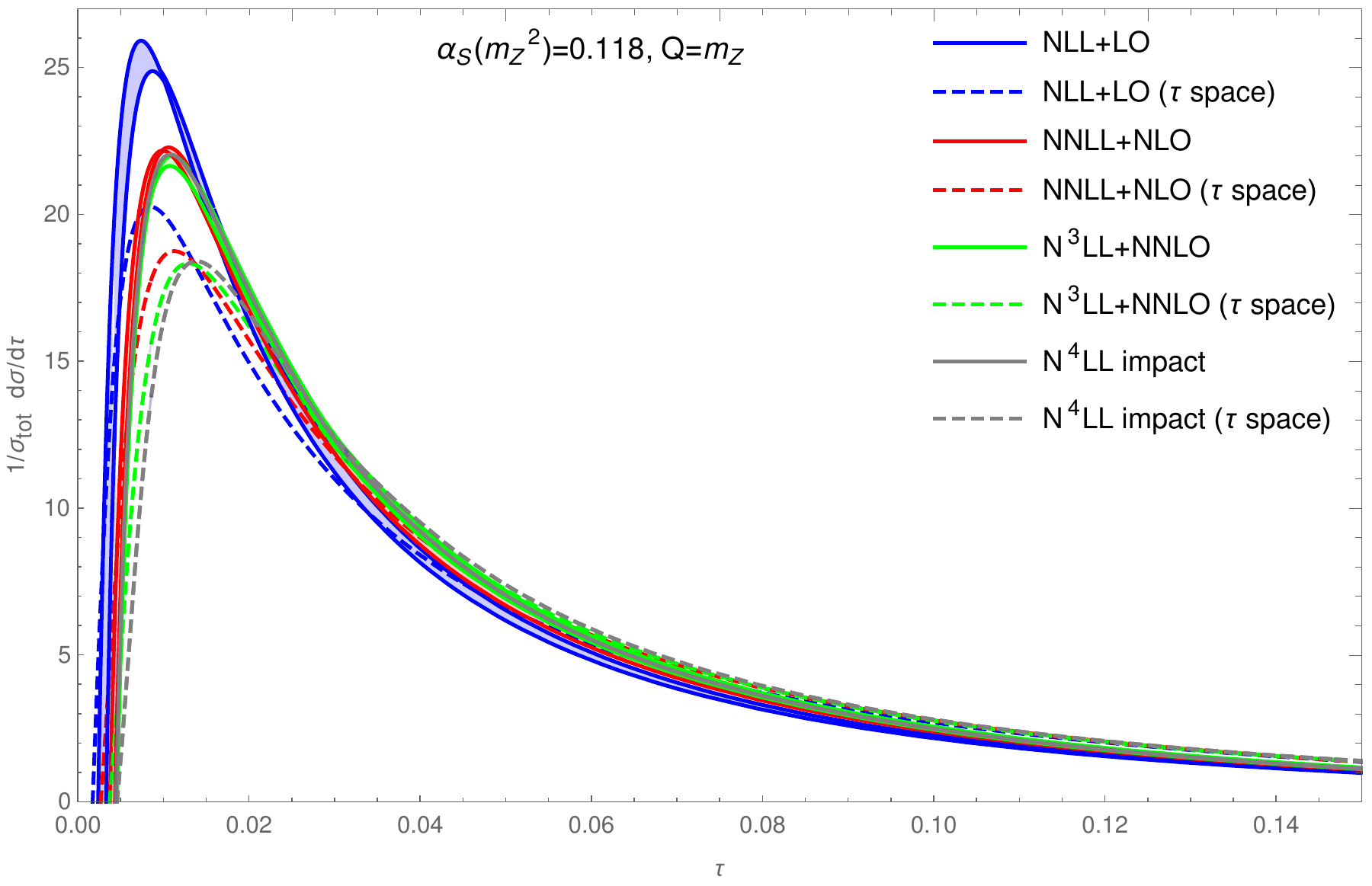}%\vspace*{-.62cm}\\
\end{center}
\caption{
\label{fig1}
{\em
The thrust distribution at $Q=91.1876$~GeV
at various perturbative orders in QCD. Results obtained through resummation in Laplace-conjugated space (solid bands) are compared with
physical $\tau$-space {\itshape approximated} results (dashed lines).
%, 
%compared with LEP data from the OPAL Collaboration.
}}
\end{figure}

In Fig.\,\ref{fig1} we show our results for the thrust distribution at $Q=m_Z=91.1876$\,GeV with $\alpha_S(m_Z^2)=0.118$.
We plot results obtained via Laplace-space resummation at NLL+LO,
NNLL+NLO and N$^3$LL+NNLO accuracies (solid bands), where uncertainty bands correspond to renormalization scale variation by a factor of two around the central value
$\mu_R=Q$. We also show the impact of N$^4$LL corrections (solid line), which is very small (differences with respect to N$^3$LL+NNLO result at few permille level). This is not unexpected because a
full next-order accuracy would also require the inclusion of the, presently unknown, $\mathcal{O}(\alpha_S^4)$ corrections both in the hard factor $C(\alpha_S)$ and in the remainder
function $D(\tau,\alpha_S)$. 

In Fig.\,\ref{fig1} we also show (dashed lines)
perturbative predictions obtained by using momentum-space resummation at NLL+LO,
NNLL+NLO, N$^3$LL+NNLO and the impact of N$^4$LL terms.
We observe that the differences among
the predictions obtained with resummation in Laplace space and $\tau$ space are substantial; these differences are larger than  the perturbative uncertainties
(estimated through both scale variation and the difference between two consecutive perturbative orders).

%%====================================
%%====================================

%%====================================
\begin{figure}[ht]
\begin{center}
\includegraphics[width=0.8\textwidth]{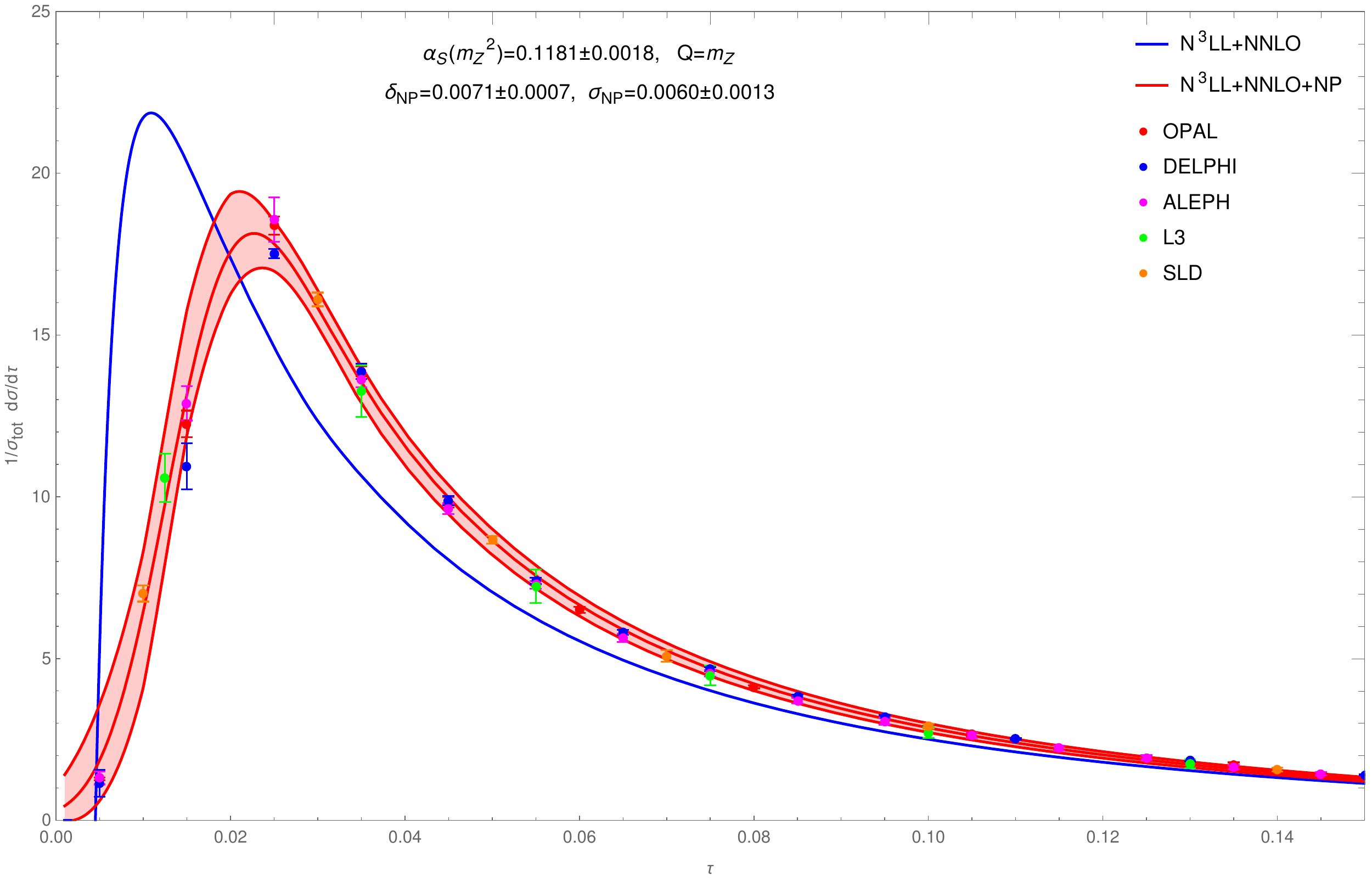}
\end{center}
\caption{
\label{fig2}
{\em
The thrust distribution at $Q=91.1876$~GeV
at N$^3$LL+NNLO in QCD without (blue solid line) and with (red band) the inclusion of NP effects. 
%, 
%compared with LEP data from the OPAL Collaboration.
}}
\end{figure}
%%====================================

Up to now, our results entirely relied on perturbation theory.
However, in order to compare our predictions with experimental data, we need to include non-perturbative (NP) hadronization effects. In this Letter, we use an analytic model based on
a correlation\,\cite{Catani:1991kz} or shape function\,\cite{Korchemsky:1999kt} $f_{\!N\!P}(\tau_h,\tau)$ depending on very few NP parameters,
such that:
\begin{equation}\label{sigNP}
\frac{d\sigma_h}{d\tau_h}=\int d\tau \frac{d\sigma}{d\tau}\,f_{\!N\!P}(\tau,\tau_h)\,,
\end{equation}
where $d\sigma_h/d\tau_h$ is the hadronic thrust distribution and $\tau_h$ is the hadronic thrust variable. We have tried various ansatz for $f_{\!N\!P}(\tau,\tau_h)$ %(\cite{})
and we have found that we can have a very good description of the
LEP and SLD data at the $Z$ boson peak ($Q=m_Z$)\,\cite{SLD:1994idb,Wicke:1999zz,DELPHI:2000uri,ALEPH:2003obs,OPAL:2004wof,L3:2004cdh}, by means of a Gaussian function depending on two free parameters only:
\begin{equation}\label{fNP}
f_{\!N\!P}(\tau_h,\tau)=\frac{1}{\sqrt{2\pi}\sigma_{\!N\!P}}\exp\left[-\frac{(\tau_h-\tau-\delta_{\!N\!P})^2}{2\,\sigma_{\!N\!P}^2}\right]\,,
\end{equation}
where $\delta_{\!N\!P}$ represents a shift and $\sigma_{\!N\!P}$ a smearing of the perturbative prediction\,\footnote{Different functions, depending on more parameters, such as an asymmetry or skewness
parameter, only marginally improve the description of the data.}.

We finally performed a three parameter ($\alpha_S(m_Z^2)$, $\delta_{\!N\!P}$, $\sigma_{\!N\!P}$) fit in the small/intermediate $\tau$  region ($0<\tau< 0.15$), obtaining, at N$^3$LL+NNLO accuracy,
the following values:
\begin{equation}
\alpha_S(m_Z^2)=0.1181\pm0.0018\,, % 0.1177\pm0.0020 with exp corr.
\qquad \delta_{\!N\!P}=0.0071 % 0.0071
\pm\,0.0007\,,%  0.0005
\qquad \sigma_{\!N\!P}=0.0060 % 0.0059
\pm0.0013\,, % 0.0011
\end{equation}
where the uncertainties include experimental and theoretical (perturbative) errors (the latter estimated by means of a renormalization scale variation of a factor two).

A comparison of our best prediction against experimental  LEP and SLD data at the $Z$ boson peak ($Q=m_Z$) is shown in Fig.(\ref{fig2}), where the uncertainty band
is obtained by a variation of the extracted parameters within the uncertainties.
As a check of our results, we have performed the fit at NNLL+NLO accuracy which gives a result (fully consistent with the N$^3$LL+NNLO one):\\
$\alpha_S(m_Z^2)=0.1194\pm 0.0020 \,,~\delta_{\!N\!P}=0.0071\pm 0.0007\,,~\sigma_{\!N\!P}=0.0062\pm 0.0014$%
%$\alpha_S(m_Z^2)=0.1188\pm 0.0021 \,,~\delta_{\!N\!P}=0.0071\pm 0.0005\,,~\sigma_{\!N\!P}=0.0061\pm 0.0012$%% with exp cprr.
\,\footnote{We made also a fit including the (known) N$^4$LL corrections, which,
as expected by the smallness of latter, gives basically the same results of the N$^3$LL+NNLO analysis
($\alpha_S(m_Z^2)=0.1177\pm 0.0018$ with $\delta_{\!N\!P}$ and $\sigma_{\!N\!P}$ unchanged).}.
% ($\alpha_S(m_Z^2)=0.1176\pm 0.0019$ with del 0.0072 0006 and sig 0.0060 0011}.
 
Conversely, an analogous fit performed using the $\tau$-space resummation formalism, gives a sensibly lower value of the strong coupling
$\alpha_S(m_Z^2)=0.1120\pm 0.0019$ %\pm0.0015$
(with $\delta_{\!N\!P}=0.0083\pm 0.0010$ and $\sigma_{\!N\!P}=0.0055\pm 0.0020$)
%$\alpha_S(m_Z^2)=0.1120\pm 0.0019$ (with $\delta_{\!N\!P}=0.0083\pm 0.0010$ and $\sigma_{\!N\!P}=0.0055\pm 0.0020$)
thus indicating that the approximations used to perform the resummation  in $\tau$ space have to be properly included in order to obtain a reliable determination of
$\alpha_S(m_Z^2)$ within a $\tau$-space resummation formalism.

In this Letter, we have presented a resummed calculation of the thrust distribution in electron-positron annihilation into hadrons up to full N$^3$LL+NNLO accuracy
(including also the N$^4$LL terms available)
in QCD perturbation theory.
We have performed the resummation directly in the Laplace-conjugated space  and we have evaluated exactly (in numerical way)
the inverse Laplace  transform.
We compared our theoretical distributions
with LEP and SLD data at the $Z$-boson mass ($m_Z$) energy, which enabled us to extract the value of the QCD coupling 
$\alpha_S(m_Z^2) = 0.1181 \pm 0.0018$,
fully consistent with the world average.
We have also shown that the commonly used (approximate) analytic  
formalism in the physical space
 gives quite different results  with respect to the Laplace-space formalism, with a corresponding lower determination of $\alpha_S(m_Z^2)$.

\vspace{1cm}
\centerline{\bf Acknowledgments}
\vspace{.2cm}
We would like to thank Stefano Camarda, Stefano Forte and Lorenzo Rossi for useful discussions on $\alpha_S$ determination and Bryan Webber for comments on the manuscript.
~\\

%\newpage
\appendix*
\section{Resummation coefficients}
In this Appendix we give the explicit expressions of the resummation coefficients up to N$^4$LL and N$^3$LO accuracy. 
The functions $f_i(\lambda)$ %for $1\leq i \leq 5$
in Eq.\,(\ref{ffun}) are:
\begingroup
\allowdisplaybreaks
\begin{footnotesize}
\begin{flalign}\label{eq:f1}
    f_1(\lambda) &=-\frac{A_1}{\beta_0 \lambda } \left[{\tilde\lambda} \ln {\tilde\lambda}-2 \bar\lambda \ln \bar\lambda\right]\,,\\
    f_2(\lambda) &= \frac{A_2}{\beta_0^2}\ln\frac{\tilde\lambda}{\bar\lambda^2}
  +\frac{2\gamma_E  A_1}{\beta_0} \ln\frac{\tilde\lambda}{\bar\lambda} %\qty[\ln {\tilde\lambda}-\ln \bar\lambda]
  +\frac{A_1 \beta_1}{\beta_0^3} \left[\ln ^2\bar\lambda-\frac{\ln ^2{\tilde\lambda}}{2} -\ln\frac{\tilde\lambda}{\bar\lambda^2} \right]%
    +\frac{B_1}{\beta_0}\ln \bar\lambda%
\,,\\
  f_3(\lambda) &= 
  -\frac{A_3 \lambda^2}{\beta_0^2 \bar\lambda {\tilde\lambda}}
  +\frac{\beta_1 A_2}{\beta_0^3 \bar\lambda {\tilde\lambda}} \left[3 \lambda ^2 + \bar\lambda \ln {\tilde\lambda}- 2 {\tilde\lambda} \ln \bar\lambda\right]
  -\frac{2\gamma_E A_2 \lambda }{ \beta_0 \bar\lambda {\tilde\lambda}}
  +\frac{\gamma_E ^2 A_1 \lambda  (2 \lambda -3)}{\bar\lambda {\tilde\lambda}}
  & \nonumber \\  &
  - \frac{2\beta_1 \gamma_E A_1}{  \beta_0^2 \bar\lambda {\tilde\lambda}}  \left[{\tilde\lambda}  \ln \bar\lambda- \bar\lambda  \ln {\tilde\lambda}-\lambda\right] 
  -\frac{\beta_1^2 A_1}{\beta_0^4} \Bigl[\frac{\lambda^2}{\bar\lambda {\tilde\lambda}}-\frac{\ln \bar\lambda}{\bar\lambda} (2 \lambda +\ln \bar\lambda)
    +\frac{\ln ^2{\tilde\lambda}}{2{\tilde\lambda}} +\frac{2 \lambda}{{\tilde\lambda}}  \ln {\tilde\lambda}\Bigr] %& \nonumber\\    &
  \nonumber \\&
  -\frac{\beta_2 A_1}{\beta_0^3 \bar\lambda {\tilde\lambda}} \left[\lambda ^2+ \bar\lambda {\tilde\lambda}(\ln {\tilde\lambda}-2 \ln \bar\lambda)\right]
- \frac{B_2\lambda}{\beta_0 \bar\lambda} + \frac{B_1 \beta_1 }{\beta_0^2 \bar\lambda}[\lambda +\ln \bar\lambda] - \frac{\gamma_E B_1 \lambda}{\bar\lambda}
\,,\\  %& \nonumber\\&
  f_4(\lambda) &=
  -\frac{A_4 \lambda ^2 \left(2 \lambda ^2-6 \lambda +3\right)}{3  \beta_0^2 {\bar\lambda}^2 {{\tilde\lambda}}^2}
  +\frac{\beta_1 A_3}{6  \beta_0^3 {\bar\lambda}^2 {{\tilde\lambda}}^2} \Bigl[15 \lambda ^2 +10 (\lambda -3) \lambda^3+3 {\bar\lambda}^2 \ln {\tilde\lambda}-6 {\tilde\lambda}^2 \ln \bar\lambda\Bigr]
  +\frac{\gamma_E A_3 \lambda  (3 \lambda -2)}{\beta_0 {\bar\lambda}^2 {{\tilde\lambda}}^2}
  & \nonumber\\  &
  +\frac{\gamma_E ^2 A_2 \lambda  \left(4 \lambda ^3-12 \lambda ^2+15 \lambda -6\right)}{{\bar\lambda}^2 {{\tilde\lambda}}^2}
  +\frac{\beta_1 \gamma_E A_2}{\beta_0^2 {\bar\lambda}^2 {{\tilde\lambda}}^2}  \Bigl[\lambda  (2-3 \lambda ) 
    +2 {\bar\lambda}^2 \ln {\tilde\lambda}-2 {\tilde\lambda}^2 \ln \bar\lambda\Bigr] +\frac{2\beta_2 A_2 \lambda ^3 (4 \lambda -3)}{3  \beta_0^3 {\bar\lambda}^2 {{\tilde\lambda}}^2}& \nonumber\\
  &-\frac{\beta_1^2 A_2}{6  \beta_0^4 {\bar\lambda}^2 {{\tilde\lambda}}^2} \Bigl[\lambda ^2 \left(22 \lambda ^2-30 \lambda +9\right)+3 {\bar\lambda}^2 \ln ^2{\tilde\lambda}
    +3 {\bar\lambda}^2 \ln {\tilde\lambda}-6 {\tilde\lambda}^2 \ln \bar\lambda (\ln \bar\lambda+1)\Bigr] 
  +\frac{\beta_1 \gamma_E ^2 A_1}{\beta_0 {\bar\lambda}^2 {{\tilde\lambda}}^2} (2 {\bar\lambda}^2 \ln {\tilde\lambda}-{\tilde\lambda}^2 \ln \bar\lambda)
  & \nonumber\\&
 +\frac{\beta_0 \gamma_E ^3 A_1 \lambda  \left(12 \lambda ^3-36 \lambda ^2+39 \lambda -14\right)}{3   {\bar\lambda}^2 {{\tilde\lambda}}^2} %&  \nonumber\\&
  -\frac{\beta_1^2 \gamma_E A_1}{\beta_0^3 {\bar\lambda}^2 {{\tilde\lambda}}^2} \left[(4 \lambda -3) \lambda ^2-{\tilde\lambda}^2 \ln ^2\bar\lambda+{\bar\lambda}^2 \ln ^2{\tilde\lambda}\right]&\nonumber\\
  &-\frac{\beta_1^3 A_1}{6   \beta_0^5 {\bar\lambda}^2 {{\tilde\lambda}}^2} \Bigl[4 (3-4 \lambda ) \lambda ^3+2 {\tilde\lambda}^2 \ln \bar\lambda \left(\ln ^2\bar\lambda-3 \lambda ^2\right)
    +12 {\bar\lambda}^2 \lambda ^2 \ln {\tilde\lambda}-{\bar\lambda}^2 \ln ^3{\tilde\lambda}\Bigr] +\frac{\beta_2 \gamma_E A_1 \lambda ^2 (4 \lambda -3)}{  \beta_0^2 {\bar\lambda}^2 {{\tilde\lambda}}^2}
  & \nonumber\\
  &+\frac{\beta_1 \beta_2 A_1}{6   \beta_0^4 {\bar\lambda}^2 {{\tilde\lambda}}^2} \Bigl[\lambda ^2 (2 \lambda  (3-7 \lambda )+3)+3 \left(8 \lambda ^2-4 \lambda +1\right) {\bar\lambda}^2 \ln {\tilde\lambda}
    -6 {\tilde\lambda}^2 (-2 \bar\lambda \lambda +1) \ln \bar\lambda\Bigr] 
  &\nonumber\\    &
  -\frac{\beta_3 A_1}{6   \beta_0^3 {\bar\lambda}^2 {{\tilde\lambda}}^2} \Bigl[(2 (\lambda -3) \lambda +3) \lambda ^2
    -6 \left(2 \lambda ^2-3 \lambda +1\right)^2 \ln \bar\lambda+3 \left(2 \lambda ^2-3 \lambda +1\right)^2 \ln {\tilde\lambda}\Bigr]
  - \frac{B_3 (2-\lambda) \lambda }{2  \beta_0 {\bar\lambda}^2}
  -\frac{\gamma_E B_2 (2-\lambda) \lambda}{{\bar\lambda}^2}
  &\nonumber\\ &
  -\frac{\beta_1 B_2 \left(\lambda ^2-2 \lambda -2 \ln \bar\lambda\right)}{2  \beta_0^2 {\bar\lambda}^2} 
  -\frac{\beta_0 \gamma_E ^2 B_1 (2-\lambda) \lambda }{2   {\bar\lambda}^2}
  +\frac{\beta_1 \gamma_E B_1 \ln \bar\lambda}{\beta_0 {\bar\lambda}^2} - \frac{\beta_2 B_1 \lambda ^2}{2 \beta_0^2 {\bar\lambda}^2}
  + \frac{\beta_1^2 B_1 (\lambda -\ln \bar\lambda) (\lambda +\ln \bar\lambda)}{2   \beta_0^3 {\bar\lambda}^2}
   \,, &\\
%%%%%%%%
  %%%%%%%%
  f_5(\lambda) &= -\frac{A_5 \lambda^2 }{6\beta_0^2 \tilde{\lambda}^3\bar{\lambda}^3} P_1(\lambda)
  -\frac{A_4 }{18 \beta_0^3 \tilde{\lambda}^3\bar{\lambda}^3} \left[ 12\lambda P_5(\lambda) \beta_0^2 \gamma_E - \beta_1 \left( 6 \bar{\lambda}^3 \ln \tilde{\lambda} - 12 \tilde{\lambda}^3 \ln \bar{\lambda} + 7\lambda^2 P_1(\lambda) \right) \right] \nonumber
  \\  &
+  \frac{A_3}{18 \beta_0^4 {\bar\lambda}^3 {\tilde\lambda}^3}
  \bigg[-9 \beta_1^2 {\bar\lambda}^3 \ln^2{\tilde\lambda}+18 \beta_1^2 {\tilde\lambda}^3 \ln^2{\bar\lambda}+6 \left(6 \gamma_E  \beta_0^2-\beta_1\right) \beta_1 {\bar\lambda}^3 \ln{\tilde\lambda}
  -12 \left(3 \gamma_E  \beta_0^2-\beta_1\right) \beta_1 {\tilde\lambda}^3 \ln{\bar\lambda}
\nonumber \\  &
  +\lambda (\lambda  \left(9 \beta_0 \beta_2 \lambda  \left(4 \lambda^3-18 \lambda^2+19\lambda -6\right)
  +\beta_1^2 \left(-52 \lambda^4+234 \lambda^3-303 \lambda^2+150\lambda -24\right)\right)+12 \gamma_E  \beta_1 \beta_0^2 P_5(\lambda )
\nonumber \\  &
  -18 \gamma_E^2 \beta_0^4 (3 P_5(\lambda)-2\lambda P_1(\lambda)))\bigg]
  +\frac{A_2}{18 \beta_0^5 {\bar\lambda}^3 {\tilde\lambda}^3}
  \Bigg[\lambda  (-12 \gamma_E  \left(\beta_0 \beta_2-\beta_1^2\right)\beta_0^2 \lambda  \left(18 \lambda^2-25 \lambda +9\right)
  \nonumber \\  &
    +\lambda(3 \beta_0^2\beta_3 \lambda  \left(20 \lambda^3-54 \lambda^2+45 \lambda -12\right)
    +\beta_1^3\left(100 \lambda^4-342 \lambda^3+339 \lambda^2-114 \lambda +6\right)
  \nonumber \\  &
    -2 \beta_0 \beta_2 \beta_1 \left(80 \lambda^4-252 \lambda^3+237 \lambda^2-75 \lambda+3\right))
    -12 \gamma_E^3 \beta_0^6 \lambda P_3(\lambda))
  \nonumber \\  &    
    +6 \beta_1^3 {\bar\lambda}^3 \ln^3{\tilde\lambda}-12 \beta_1^3 {\tilde\lambda}^3 \ln^3{\bar\lambda}
    -36\gamma_E  \beta_0^2 \beta_1^2 ({\bar\lambda}^3 \ln^2{\tilde\lambda}- {\tilde\lambda}^3 \ln^2{\bar\lambda})
  \nonumber \\  &    
    +6 \beta_1 \left(12\gamma_E^2 \beta_0^4-\left(\beta_1^2-\beta_0 \beta_2\right) (6 \lambda -1)\right){\bar\lambda}^3 \ln{\tilde\lambda}
    -12 \beta_1 \left(3 \gamma_E^2 \beta_0^4-\left(\beta_1^2-\beta_0 \beta_2\right) (3 \lambda -1)\right) {\tilde\lambda}^3 \ln{\bar\lambda}\Bigg] 
  \nonumber \\  &
  +\frac{2 A_1}{{\bar\lambda}^3 {\tilde\lambda}^3}
  \Bigg[\frac{\beta_1^4 }{24 \beta_0^6}
    \Bigg(-56 \lambda^6+108 \lambda^5-66 \lambda^4+12 \lambda^3-8 (4 \lambda -3) \bar\lambda^3 \lambda^2 \ln\tilde\lambda +4 (2 \lambda -3)
    \tilde\lambda^3 \lambda^2 \ln\bar\lambda
    \nonumber \\ &
+12 \bar\lambda^3 \lambda \ln^2\tilde\lambda    -12 \tilde\lambda^3 \lambda \ln^2\bar\lambda  -\bar\lambda^3 \ln^4\tilde\lambda
    +2 \bar\lambda^3 \ln^3\tilde\lambda +2\tilde\lambda^3 \ln^4\bar\lambda -4 {\tilde\lambda}^3 \ln^3\bar\lambda \Bigg) \nonumber \\ &
    +\frac{\gamma_E\beta_1^3}{6 \beta_0^4}  \left(2 \bar\lambda^3\ln^3{\tilde\lambda} -3 \bar\lambda^3\ln^2\tilde\lambda -12 \lambda \bar\lambda^3 \ln\tilde\lambda -2 \tilde\lambda^3\ln^3\bar\lambda
      +3 \tilde\lambda^3 \ln^2\bar\lambda +6 \lambda {\tilde\lambda}^3 \ln\bar\lambda +\lambda^2 P_2(\lambda)\right) 
    \nonumber \\ &
    +\frac{\beta_1^2\gamma_E^2}{2 \beta_0^2} \left(4 \lambda^4-6 \lambda^2+3 \lambda -2 {\bar\lambda}^3\ln^2\tilde\lambda +2 \bar\lambda^3\ln\tilde\lambda +{\tilde\lambda}^3\ln^2\bar\lambda -\tilde\lambda^3\ln\bar\lambda \right)
    \nonumber \\ &    
    -\frac{\beta_1^2\beta_2}{36 \beta_0^5}\Bigg(-164 \lambda^6+306 \lambda^5-165 \lambda^4+12 \lambda^3+6 \lambda^2+18 \bar\lambda^3 \lambda\ln^2\tilde\lambda  -18 {\tilde\lambda}^3 \lambda\ln^2\bar\lambda
    \nonumber \\ &    
    +12 \ln\bar\lambda \left(3 \lambda  \bar\lambda^2-1\right)\tilde\lambda^3+\left(144 \lambda^6-576 \lambda^5+900 \lambda^4-690 \lambda^3+270\lambda^2-54 \lambda +6\right) \ln\tilde\lambda\Bigg)    
    \nonumber \\ &
    +\frac{\beta_1\gamma_E^3}{6}  \left(-\lambda P_3(\lambda) +8 \bar\lambda^3\ln\tilde\lambda-2 \tilde\lambda^3\ln\bar\lambda \right)
    \nonumber \\ &
    -\frac{\gamma_E\beta_1   \beta_2 \lambda}{3 \beta_0^3} \left(-6 \bar\lambda^3\ln\tilde\lambda +3 \tilde\lambda^3\ln{\bar\lambda}       +\lambda  P_2(\lambda )\right)
    +\frac{\beta_1\beta_3}{18 \beta_0^4}\bigg(-28 \lambda^6+72 \lambda^5-51 \lambda^4+6 \lambda^3+3 \lambda^2
   \nonumber \\ &
    -3 (2 \lambda (\lambda  (8 \lambda -9)+3)-1) \bar\lambda^3\ln\tilde\lambda
     +3 (\lambda (\lambda  (4 \lambda -9)+6)-2) \tilde\lambda^3 \ln\bar\lambda  \bigg)
     +\frac{\beta_0^2\gamma_E^4}{12}  \lambda  (2 \lambda -3) P_4(\lambda )
   \nonumber \\ &
    +\frac{\gamma_E  \lambda^2  \beta_3}{6 \beta_0^2}P_2(\lambda )
    +\frac{\beta_2^2}{36 \beta_0^4} \left(-20 \lambda^6-18 \lambda^5+69 \lambda^4-42 \lambda^3+6\lambda^2-12 \bar\lambda^3 \tilde\lambda^3\ln\bar\lambda +6 \bar\lambda^3 \tilde\lambda^3\ln{\tilde\lambda} \right)
   \nonumber \\ &
   -\frac{\beta_4}{36 \beta_0^3} \left(\lambda^2P_1(\lambda) -12 \bar\lambda^3 \tilde\lambda^3 \ln\bar\lambda+6 \bar\lambda^3 {\tilde\lambda}^3 \ln\tilde\lambda \right)
     -\frac{\gamma_E^2 \beta_2}{2\beta_0}\lambda  \left(4 \lambda^3-6 \lambda +3\right)
   \Bigg]
\nonumber \\  
  & -\frac{B_4 \lambda}{3 \beta_0 \bar{\lambda}^3} P_6(\lambda) + \frac{B_3}{3 \beta_0^2 \bar{\lambda}^3} \left[ (\lambda P_6(\lambda) + 3 \ln \bar{\lambda}) \beta_1 - 3\lambda P_6(\lambda) \beta_0^2 \gamma_E \right] \nonumber \\
& -\frac{B_2}{3 \beta_0^3 \bar{\lambda}^3} \left[ 3\lambda P_6(\lambda) \gamma_E^2 \beta_0^4 - 6\ln \bar{\lambda} \beta_1 \gamma_E \beta_0^2 - (\lambda -3) \lambda^2 \beta_2 \beta_0 + \left( (\lambda-3)\lambda^2 + 3\ln^2 \bar{\lambda} \right) \beta_1^2 \right] \nonumber \\
& -\frac{B_1}{6 \beta_0^4 \bar{\lambda}^3} \Bigg[ 2\lambda P_6(\lambda) \gamma_E^3 \beta_0^6 + 3 \left( \lambda P_6(\lambda) - 2\ln \bar{\lambda} \right) \beta_1 \gamma_E^2 \beta_0^4 + 6\lambda \beta_2 \gamma_E \beta_0^3 \nonumber \\
& + \left( (3-2\lambda) \lambda^2 \beta_3 - 6(-\ln^2 \bar{\lambda} + \ln \bar{\lambda} + \lambda) \beta_1^2 \gamma_E \right) \beta_0^2 + 2\lambda \left( \lambda(2\lambda-3) - 3\ln \bar{\lambda} \right) \beta_1 \beta_2 \beta_0 \nonumber \\
    & + \left( -2\ln^3 \bar{\lambda} + 3\ln^2 \bar{\lambda} + 6\lambda \ln \bar{\lambda} + (3-2\lambda) \lambda^2 \right) \beta_1^3 \Bigg]\,,\\
  %%%%%%%%%
P_1(\lambda) &= 4\lambda^4 - 18\lambda^3 + 33\lambda^2 - 24\lambda + 6\,,\qquad
P_2(\lambda) = 12 \lambda^3 - 36 \lambda^2 + 32 \lambda  -9 \,, \nonumber \\
P_3(\lambda) &= -24\lambda^5 + 108\lambda^4 - 198\lambda^3 + 193\lambda^2 - 99\lambda + 21\,,
\qquad P_4(\lambda) = 28\lambda^4 - 84\lambda^3 + 105\lambda^2 - 63\lambda + 15\,, \nonumber \\
P_5(\lambda) &= 7\lambda^2 - 9\lambda + 3\,,\qquad
P_6(\lambda) = \lambda^2 - 3\lambda + 3\,,
\end{flalign}
\end{footnotesize}
\endgroup
\noindent where $\lambda =  \frac{\beta_0}{\pi}\alpha_s \ln N$, $\bar\lambda=1-\lambda$, $\tilde\lambda=1-2\lambda$, %,  $\alpha_s = \alpha_s(\mu_R^2)$, $l_R=\ln(Q^2/\mu_R^2)$ and
and $\gamma_E=0.5772\cdots$ is the Euler-Mascheroni constant.
The QCD $\beta$-function coefficients read, up to five loops\,\cite{Tarasov:1980au,Larin:1993tp,vanRitbergen:1997va,Czakon:2004bu,Herzog:2017ohr}:
\begin{footnotesize}
\begin{flalign}
\beta_0 &= \frac{11 C_A \, - \, 2 n_f}{12}\,, % \, = \, \frac{33 \, - \, 2 n_f}{12}\,
\qquad\qquad \beta_1 = \frac{1}{24} \left( 17 C_A^2 - 5 C_A n_f - 3 C_F n_f \right)\,, 
\\
\beta_2 &=  \frac{1}{64} 
\left[
\frac{2857}{54} \, C_A^3
\, - \, 
\left(
\frac{1415}{54} C_A^2 \, + \, \frac{205}{18} C_A C_F \, - \, C_F^2
\right) n_f
\, + \,
\left(
\frac{79}{54} C_A + \frac{11}{9} C_F 
\right) n_f^2
\right]\,,
\\
\beta_3 &=
\frac{1093}{186624}n_f^3 + n_f^2 \left(\frac{809 \zeta_3}{2592}
+ \frac{50065}{41472}\right) + n_f
   \left( - \frac{1627 \zeta_3}{1728} - \frac{1078361}{41472} \right)
   +\frac{891}{64} \zeta_3 + \frac{149753}{1536}\,,
\\
\beta_4 &= -0.00179929\, n_f^4-0.225857\, n_f^3+17.156\, n_f^2-181.799\,
   n_f+524.558\,,
\end{flalign}
\end{footnotesize}
where $n_f$ is the number of QCD active
(effective massless) flavors at the hard scale $Q$ ($n_f=5$ for $Q=m_Z$) and $\zeta_3=1.20206\cdots$.
The coefficients $A_i$ and $B_i$ are extracted from the fixed-order calculations for the thrust distribution~\cite{Becher:2008cf, Bruser:2018rad,  Kelley:2011ng,Chen:2020dpk, Baranowski:2021gxe, Baranowski:2022khd,Moult:2022xzt,Duhr:2022yyp,Duhr:2022cob,Baranowski:2024ysi,Baranowski:2024vxg} and the cusp anomalous dimensions~\cite{Moch:2004pa, Henn:2019swt, vonManteuffel:2020vjv, Herzog:2018kwj}:
\begin{footnotesize}
\begin{align}
      A_1 &=C_F \,,\qquad\qquad    A_2 = \frac{C_F}{2}  \left[\left(\frac{67}{18}-\frac{\pi^2}{6}\right) C_A-\frac59 n_f\right]\,,\\
      A_3 &= \frac{25+3 \pi^2}{243} n_f^2+\frac{288 \zeta_3-72 \pi ^2-2381}{324} n_f%\nonumber\\&
      +\frac{77515-390 \pi ^2-35640 \zeta_3+198 \pi ^4}{1080}\,,\\
   A_4&= \left(\frac{4 \zeta_3}{81}-\frac{\pi ^2}{27}+\frac{91}{4374}\right)n_f^3+ \left(-\frac{407 \zeta_3}{108}-\frac{31 \pi ^4}{1080}+\frac{23 \pi^2}{12}+\frac{147199}{46656}\right)n_f^2\nonumber\\
   &+ \left(-\frac{125 \zeta_5}{18}-\frac{7 \pi ^2 \zeta_3}{36}+\frac{71057 \zeta_3}{648}+\frac{143 \pi ^4}{180}-\frac{266 \pi^2}{9}-\frac{1820789}{15552}\right)n_f\nonumber\\
   &+\frac{3025 \zeta_5}{12}-6 \zeta_3^2+\frac{99 \pi ^2 \zeta_3}{8}-\frac{24451 \zeta_3}{24}-\frac{2033 \pi ^6}{15120}-\frac{759 \pi^4}{160}+\frac{20483 \pi ^2}{144}+\frac{4311229}{5184}\,,\\
   A_5&= 0.001302 \, \Gamma_{\mathrm{cusp}}^{(4)} + 16445.4 - 5225.93\, n_f + 567.497\, n_f^2 - 24.5952\, n_f^3 + 0.356813\, n_f^4\,,\\
%         0.001302 \, \Gamma_{\mathrm{cusp}}^{(4)}-0.6888  \, c_{3,qq}+708.4\,\simeq\, 1716.7\qquad (\mbox{for~} n_f=5)\,,\\
%\end{align}
%and
%\begin{align}
      B_1&=-\frac{3 C_F}{2}\,,\qquad
       B_2=C_A C_F\left(\frac{3  \zeta_3}{2}-\frac{57}{16}+\frac{11\pi^2}{24} \right)+C_F^2\left(-3 \zeta_3+\frac{\pi^2}{4}-\frac{3}{16}\right)+C_F n_f\left(\frac{5 }{8} -\frac{\pi^2}{12} \right)\,,\\
   B_3&=\left(-\frac{20  \zeta_3}{81}+\frac{173 \pi^2}{972}-\frac{191}{324}\right)n_f^2+\left(\frac{62\zeta_3}{9}+\frac{52\pi^4}{405}-\frac{1109 \pi^2}{162}+\frac{2467}{108}\right) n_f\nonumber\\
   &-\frac{155 \zeta_5}{9}-\frac{4 \pi ^2 \zeta_3}{81}-\frac{481 \zeta_3}{27}-\frac{2539 \pi ^4}{1080}+\frac{26189 \pi ^2}{432}-\frac{18797}{108}\,,\\
     B_4&= 2592.79 - 620.792\, n_f + 44.0707\, n_f^2 - 0.949394\, n_f^3\,, % 225.8 -0.1797 \,c_{3,qq}\simeq 471.9\qquad (\mbox{for~}  n_f=5)\,,
\end{align}
\end{footnotesize}
where $\zeta_5=1.03693\cdots$ and %$c_{3,qq}= -1369.5758$ \cite{Baranowski:2024vxg} and
$\Gamma_{\mathrm{cusp}}^{(4)}\simeq 49873.5$ for $n_f=5$ \cite{Herzog:2018kwj}.
The hard-virtual coefficients $C_i$ are derived from the quark form factor \cite{Baikov:2009bg,Lee:2010cga,Gehrmann:2010ue,Lee:2022nhh} and the constant terms in the collinear/soft functions \cite{Becher:2008cf, Bruser:2018rad,Kelley:2011ng,Chen:2020dpk, Baranowski:2021gxe, Baranowski:2022khd,Moult:2022xzt,Duhr:2022yyp,Duhr:2022cob,Baranowski:2024ysi,Baranowski:2024vxg}%
\footnote{The numerical impact of the, usually neglected, singlet diagrams induced by axial-vector currents and their implementation, can be found in  \cite{Ju:2021lah}.}:
\begin{footnotesize}
\begin{flalign}
       C_1 &= C_F \left(-\frac{5}{4}+\frac{3 \gamma_E }{2}-\gamma_E ^2\right)\,,\\
       C_2 &=C_AC_F\left(\frac{41 \zeta_3}{6}-\frac{3 \gamma_E  \zeta_3}{2}-\frac{491}{96}+\frac{57 \gamma_E }{16}-\frac{169 \gamma_E ^2}{144}-\frac{11 \gamma ^3}{12}-\frac{275 \pi ^2}{864}-\frac{11
   \gamma_E  \pi ^2}{24}+\frac{\gamma_E ^2 \pi ^2}{12}+\frac{31 \pi ^4}{1440}\right)\nonumber\\
   &+C_F^2 \left(-\frac{9 \zeta_3}{2}+3 \gamma_E  \zeta_3+\frac{41}{32}-\frac{27 \gamma_E }{16}+\frac{19 \gamma_E
   ^2}{8}-\frac{3 \gamma _E^3}{2}+\frac{\gamma_E ^4}{2}+\frac{17 \pi ^2}{96}-\frac{\gamma_E  \pi ^2}{4}+\frac{19 \pi ^4}{720}\right)\nonumber\\
   &+C_Fn_f\left(-\frac{5 \zeta_3}{6}+\frac{35}{48}-\frac{5
   \gamma_E }{8}+\frac{11 \gamma_E ^2}{72}+\frac{\gamma_E ^3}{6}+\frac{25 \pi ^2}{432}+\frac{\gamma _E \pi ^2}{12}\right)\,,\\
       C_3 &= -0.573462\, n_f^2+4.90384\, n_f+13.4156\,. %0.015625\, c_S^{(3)}+44.998\,.
\end{flalign}
\end{footnotesize}
To avoid cumbersome formulae, in the higher-order terms %$f_5(\lambda)$, $\beta_3$, $\beta_4$, $A_3$, $A_4$, $A_5$, $B_3$, $B_4$, $C_2$ and $C_2$
we have replaced the explicit value of the color factors of $SU(N_c=3)$, namely $C_F = 4/3$, $C_A = 3$.

\noindent

%%%%%%%%%%%%%%%%%%%%%%%%%%%


\begin{thebibliography}{99}

%\cite{Farhi:1977sg}
\bibitem{Farhi:1977sg}
E.~Farhi,
%``A QCD Test for Jets,''
Phys. Rev. Lett. \textbf{39} (1977), 1587-1588
doi:10.1103/PhysRevLett.39.1587
%1142 citations counted in INSPIRE as of 21 Jan 2025

%\cite{Gehrmann-DeRidder:2007nzq}
\bibitem{Gehrmann-DeRidder:2007nzq}
A.~Gehrmann-De Ridder, T.~Gehrmann, E.~W.~N.~Glover and G.~Heinrich,
%``Second-order QCD corrections to the thrust distribution,''
Phys. Rev. Lett. \textbf{99} (2007), 132002
doi:10.1103/PhysRevLett.99.132002
[arXiv:0707.1285 [hep-ph]].
%193 citations counted in INSPIRE as of 20 Dec 2024

%\cite{Weinzierl:2008iv}
\bibitem{Weinzierl:2008iv}
S.~Weinzierl,
%``NNLO corrections to 3-jet observables in electron-positron annihilation,''
Phys. Rev. Lett. \textbf{101} (2008), 162001
doi:10.1103/ PhysRevLett.101.162001
[arXiv:0807.3241 [hep-ph]].
%196 citations counted in INSPIRE as of 20 Dec 2024

%\cite{Gehrmann-DeRidder:2009fgd}
\bibitem{Gehrmann-DeRidder:2009fgd}
A.~Gehrmann-De Ridder, T.~Gehrmann, E.~W.~N.~Glover and G.~Heinrich,
%``NNLO moments of event shapes in e+e- annihilation,''
JHEP \textbf{05} (2009), 106
doi:10.1088/1126-6708/2009/05/106
[arXiv:0903.4658 [hep-ph]].
%95 citations counted in INSPIRE as of 23 Dec 2024

%\cite{Weinzierl:2009ms}
\bibitem{Weinzierl:2009ms}
S.~Weinzierl,
%``Event shapes and jet rates in electron-positron annihilation at NNLO,''
JHEP \textbf{06} (2009), 041
doi:10.1088/1126-6708/2009/06/041
[arXiv:0904.1077 [hep-ph]].
%184 citations counted in INSPIRE as of 21 Jan 2025

%\cite{Gehrmann-DeRidder:2014hxk}
\bibitem{Gehrmann-DeRidder:2014hxk}
A.~Gehrmann-De Ridder, T.~Gehrmann, E.~W.~N.~Glover and G.~Heinrich,
%``EERAD3: Event shapes and jet rates in electron-positron annihilation at order $\alpha_s^3$,''
Comput. Phys. Commun. \textbf{185} (2014), 3331
doi:10.1016/j.cpc.2014.07.024
[arXiv:1402.4140 [hep-ph]].
%87 citations counted in INSPIRE as of 20 Dec 2024

%\cite{DelDuca:2016csb}
\bibitem{DelDuca:2016csb}
V.~Del Duca, C.~Duhr, A.~Kardos, G.~Somogyi and Z.~Tr\'ocs\'anyi,
%``Three-Jet Production in Electron-Positron Collisions at Next-to-Next-to-Leading Order Accuracy,''
Phys. Rev. Lett. \textbf{117} (2016) no.15, 152004
doi:10.1103/PhysRevLett.117.152004
[arXiv:1603.08927 [hep-ph]].
%143 citations counted in INSPIRE as of 29 Jan 2025

%\cite{DelDuca:2016ily}
\bibitem{DelDuca:2016ily}
V.~Del Duca, C.~Duhr, A.~Kardos, G.~Somogyi, Z.~Sz\H{o}r, Z.~Tr\'ocs\'anyi and Z.~Tulip\'ant,
%``Jet production in the CoLoRFulNNLO method: event shapes in electron-positron collisions,''
Phys. Rev. D \textbf{94} (2016) no.7, 074019
doi:10.1103/PhysRevD.94.074019
[arXiv:1606.03453 [hep-ph]].
%176 citations counted in INSPIRE as of 06 Jan 2025

%\cite{Catani:1991kz}
\bibitem{Catani:1991kz}
S.~Catani, G.~Turnock, B.~R.~Webber and L.~Trentadue,
%``Thrust distribution in e+ e- annihilation,''
Phys. Lett. B \textbf{263} (1991), 491-497
doi:10.1016/0370-2693(91)90494-B
%294 citations counted in INSPIRE as of 21 Jan 2025

%\cite{Catani:1992ua}
\bibitem{Catani:1992ua}
S.~Catani, L.~Trentadue, G.~Turnock and B.~R.~Webber,
%``Resummation of large logarithms in e+ e- event shape distributions,''
Nucl. Phys. B \textbf{407} (1993), 3-42
doi:10.1016/0550-3213(93)90271-P
%596 citations counted in INSPIRE as of 17 Jan 2025

%\cite{Schwartz:2007ib}
\bibitem{Schwartz:2007ib}
M.~D.~Schwartz,
%``Resummation and NLO matching of event shapes with effective field theory,''
Phys. Rev. D \textbf{77} (2008), 014026
doi:10.1103/PhysRevD.77.014026
[arXiv:0709.2709 [hep-ph]].
%154 citations counted in INSPIRE as of 20 Dec 2024

%\cite{Becher:2008cf}
\bibitem{Becher:2008cf}
T.~Becher and M.~D.~Schwartz,
%``A precise determination of $\alpha_s$ from LEP thrust data using effective field theory,''
JHEP \textbf{07} (2008), 034
doi:10.1088/1126-6708/2008/07/034
[arXiv:0803.0342 [hep-ph]].
%338 citations counted in INSPIRE as of 20 Dec 2024

%\cite{Dissertori:2009ik}
\bibitem{Dissertori:2009ik}
G.~Dissertori, A.~Gehrmann-De Ridder, T.~Gehrmann, E.~W.~N.~Glover, G.~Heinrich, G.~Luisoni and H.~Stenzel,
%``Determination of the strong coupling constant using matched NNLO+NLLA predictions for hadronic event shapes in e+e- annihilations,''
JHEP \textbf{08} (2009), 036
doi:10.1088/1126-6708/2009/08/036
[arXiv:0906.3436 [hep-ph]].
%142 citations counted in INSPIRE as of 31 Dec 2024

\bibitem{Hornig:2009vb}
A.~Hornig, C.~Lee and G.~Ovanesyan,
%``Effective Predictions of Event Shapes: Factorized, Resummed, and Gapped Angularity Distributions,''
JHEP \textbf{05} (2009), 122
doi:10.1088/1126-6708/2009/05/122
[arXiv:0901.3780 [hep-ph]].

%\cite{Abbate:2010xh}
\bibitem{Abbate:2010xh}
R.~Abbate, M.~Fickinger, A.~H.~Hoang, V.~Mateu and I.~W.~Stewart,
%``Thrust at $N^{3}LL$ with Power Corrections and a Precision Global Fit for $\alpha_{s}(mZ)$,''
Phys. Rev. D \textbf{83} (2011), 074021
doi:10.1103/PhysRevD.83.074021
[arXiv:1006.3080 [hep-ph]].
%417 citations counted in INSPIRE as of 14 Jan 2025

%\cite{Monni:2011gb}
\bibitem{Monni:2011gb}
P.~F.~Monni, T.~Gehrmann and G.~Luisoni,
%``Two-Loop Soft Corrections and Resummation of the Thrust Distribution in the Dijet Region,''
JHEP \textbf{08} (2011), 010
doi:10.1007/JHEP08(2011)010
[arXiv:1105.4560 [hep-ph]].
%159 citations counted in INSPIRE as of 20 Dec 2024

%\cite{Abbate:2012jh}
\bibitem{Abbate:2012jh}
R.~Abbate, M.~Fickinger, A.~H.~Hoang, V.~Mateu and I.~W.~Stewart,
%``Precision Thrust Cumulant Moments at $N^3$LL,''
Phys. Rev. D \textbf{86} (2012), 094002
doi:10.1103/PhysRevD.86.094002
[arXiv:1204.5746 [hep-ph]].
%97 citations counted in INSPIRE as of 21 Jan 2025

\bibitem{Almeida:2014uva}
L.~G.~Almeida, S.~D.~Ellis, C.~Lee, G.~Sterman, I.~Sung and J.~R.~Walsh,
%``Comparing and counting logs in direct and effective methods of QCD resummation,''
JHEP \textbf{04} (2014), 174
doi:10.1007/JHEP04(2014)174
[arXiv:1401.4460 [hep-ph]].

%\cite{Banfi:2014sua}
\bibitem{Banfi:2014sua}
A.~Banfi, H.~McAslan, P.~F.~Monni and G.~Zanderighi,
%``A general method for the resummation of event-shape distributions in $e^{+} e^{−}$ annihilation,''
JHEP \textbf{05} (2015), 102
doi:10.1007/JHEP05(2015)102
[arXiv:1412.2126 [hep-ph]].
%158 citations counted in INSPIRE as of 23 Dec 2024

%\cite{Baron:2018nfz}
\bibitem{Baron:2018nfz}
J.~Baron, S.~Marzani and V.~Theeuwes,
%``Soft-Drop Thrust,''
JHEP \textbf{08} (2018), 105
[erratum: JHEP \textbf{05} (2019), 056]
doi:10.1007/JHEP08(2018)105
[arXiv:1803.04719 [hep-ph]].
%44 citations counted in INSPIRE as of 15 Jan 2025

%\cite{Wang:2019ljl}
\bibitem{Wang:2019ljl}
S.~Q.~Wang, S.~J.~Brodsky, X.~G.~Wu and L.~Di Giustino,
%``Thrust Distribution in Electron-Positron Annihilation using the Principle of Maximum Conformality,''
Phys. Rev. D \textbf{99} (2019) no.11, 114020
doi:10.1103/PhysRevD.99.114020
[arXiv:1902.01984 [hep-ph]].
%13 citations counted in INSPIRE as of 15 Jan 2025

%\cite{Marzani:2019evv}
\bibitem{Marzani:2019evv}
S.~Marzani, D.~Reichelt, S.~Schumann, G.~Soyez and V.~Theeuwes,
%``Fitting the Strong Coupling Constant with Soft-Drop Thrust,''
JHEP \textbf{11} (2019), 179
doi:10.1007/JHEP11(2019)179
[arXiv:1906.10504 [hep-ph]].
%51 citations counted in INSPIRE as of 12 Jan 2025

%\cite{Benitez:2024nav}
\bibitem{Benitez:2024nav}
M.~A.~Benitez, A.~H.~Hoang, V.~Mateu, I.~W.~Stewart and G.~Vita,
%``On Determining $\alpha_s(m_Z)$ from Dijets in $e^+e^-$ Thrust,''
[arXiv:2412.15164 [hep-ph]].
%1 citations counted in INSPIRE as of 29 Jan 2025

%\cite{Gorishnii:1990vf}
\bibitem{Gorishnii:1990vf}
S.~G.~Gorishnii, A.~L.~Kataev and S.~A.~Larin,
%``The $O(\alpha^{3}_{s})$-corrections to $\sigma_{tot}(e^{+}e^{-}\rightarrow hadrons)$ and $\Gamma(\tau^{-} \rightarrow \nu_{\tau} + hadrons)$ in QCD,''
Phys. Lett. B \textbf{259} (1991), 144-150
doi:10.1016/0370-2693(91)90149-K
%819 citations counted in INSPIRE as of 08 Nov 2024

%\cite{Surguladze:1990tg}
\bibitem{Surguladze:1990tg}
L.~R.~Surguladze and M.~A.~Samuel,
%``Total hadronic cross-section in e+ e- annihilation at the four loop level of perturbative QCD,''
Phys. Rev. Lett. \textbf{66} (1991), 560-563
[erratum: Phys. Rev. Lett. \textbf{66} (1991), 2416]
doi:10.1103/PhysRevLett.66.560
%713 citations counted in INSPIRE as of 08 Nov 2024

%%%% NP effects

\bibitem{Webber:1994cp}
B.~R.~Webber,
%``Estimation of power corrections to hadronic event shapes,''
Phys. Lett. B \textbf{339} (1994), 148-150
doi:10.1016/0370-2693(94)91147-9
[arXiv:hep-ph/9408222 [hep-ph]].

\bibitem{Dokshitzer:1995zt}
Y.~L.~Dokshitzer and B.~R.~Webber,
%``Calculation of power corrections to hadronic event shapes,''
Phys. Lett. B \textbf{352} (1995), 451-455
doi:10.1016/0370-2693(95)00548-Y
[arXiv:hep-ph/9504219 [hep-ph]].

\bibitem{Nason:1995np}
P.~Nason and M.~H.~Seymour,
%``Infrared renormalons and power suppressed effects in e+ e- jet events,''
Nucl. Phys. B \textbf{454} (1995), 291-312
doi:10.1016/0550-3213(95)00461-Z
[arXiv:hep-ph/9506317 [hep-ph]].

\bibitem{Dokshitzer:1997ew}
Y.~L.~Dokshitzer and B.~R.~Webber,
%``Power corrections to event shape distributions,''
Phys. Lett. B \textbf{404} (1997), 321-327
doi:10.1016/S0370-2693(97)00573-X
[arXiv:hep-ph/9704298 [hep-ph]].

\bibitem{Gardi:1999dq}
E.~Gardi and G.~Grunberg,
%``Power corrections in the single dressed gluon approximation: The Average thrust as a case study,''
JHEP \textbf{11} (1999), 016
doi:10.1088/1126-6708/1999/11/016
[arXiv:hep-ph/9908458 [hep-ph]].

\bibitem{Gardi:2000yh}  
E.~Gardi,
%``Perturbative and nonperturbative aspects of moments of the thrust distribution in e+ e- annihilation,''
JHEP \textbf{04} (2000), 030
doi:10.1088/1126-6708/2000/04/030
[arXiv:hep-ph/0003179 [hep-ph]].

\bibitem{Gardi:2001ny}
E.~Gardi and J.~Rathsman,
%``Renormalon resummation and exponentiation of soft and collinear gluon radiation in the thrust distribution,''
Nucl. Phys. B \textbf{609} (2001), 123-182
doi:10.1016/S0550-3213(01)00284-X
[arXiv:hep-ph/0103217 [hep-ph]].

\bibitem{Lee:2006nr}
C.~Lee and G.~F.~Sterman,
%``Momentum Flow Correlations from Event Shapes: Factorized Soft Gluons and Soft-Collinear Effective Theory,''
Phys. Rev. D \textbf{75} (2007), 014022
doi:10.1103/PhysRevD.75.014022
[arXiv:hep-ph/0611061 [hep-ph]].

%\cite{Davison:2009wzs}
\bibitem{Davison:2009wzs}
R.~A.~Davison and B.~R.~Webber,
%``Non-Perturbative Contribution to the Thrust Distribution in e+ e- Annihilation,''
Eur. Phys. J. C \textbf{59} (2009), 13-25
doi:10.1140/epjc/s10052-008-0836-7
[arXiv:0809.3326 [hep-ph]].
%85 citations counted in INSPIRE as of 20 Dec 2024

%%%%%

%\cite{ParticleDataGroup:2024cfk}
\bibitem{ParticleDataGroup:2024cfk}
S.~Navas \textit{et al.} [Particle Data Group],
%``Review of particle physics,''
Phys. Rev. D \textbf{110} (2024) no.3, 030001
doi:10.1103/PhysRevD.110.030001
%852 citations counted in INSPIRE as of 29 Jan 2025

%\cite{Gehrmann:2012sc}
\bibitem{Gehrmann:2012sc}
T.~Gehrmann, G.~Luisoni and P.~F.~Monni,
%``Power corrections in the dispersive model for a determination of the strong coupling constant from the thrust distribution,''
Eur. Phys. J. C \textbf{73} (2013) no.1, 2265
doi:10.1140/epjc/s10052-012-2265-x
[arXiv:1210.6945 [hep-ph]].
%97 citations counted in INSPIRE as of 21 Jan 2025

%\cite{Nason:2023asn}
\bibitem{Nason:2023asn}
P.~Nason and G.~Zanderighi,
%``Fits of \ensuremath{\alpha}$_{s}$ using power corrections in the three-jet region,''
JHEP \textbf{06} (2023), 058
doi:10.1007/JHEP06(2023)058
[arXiv:2301.03607 [hep-ph]].
%22 citations counted in INSPIRE as of 29 Jan 2025

\bibitem{Nason:2025qbx}
P.~Nason and G.~Zanderighi,
%``Fits of $\alpha_s$ from event-shapes in the three-jet region: extension to all energies,''
[arXiv:2501.18173 [hep-ph]].

%\cite{Catani:1996yz}
\bibitem{Catani:1996yz}
S.~Catani, M.~L.~Mangano, P.~Nason and L.~Trentadue,
%``The Resummation of soft gluons in hadronic collisions,''
Nucl. Phys. B \textbf{478} (1996), 273-310
doi:10.1016/0550-3213(96)00399-9
[arXiv:hep-ph/9604351 [hep-ph]].
%554 citations counted in INSPIRE as of 24 Jan 2025

%\cite{Aglietti:2002ew}
\bibitem{Aglietti:2002ew}
U.~Aglietti and G.~Ricciardi,
%``Approximate NNLO threshold resummation in heavy flavor decays,''
Phys. Rev. D \textbf{66} (2002), 074003
doi:10.1103/PhysRevD.66.074003
[arXiv:hep-ph/0204125 [hep-ph]].
%23 citations counted in INSPIRE as of 25 Sep 2024


%\cite{Catani:1997xc}
\bibitem{Catani:1997xc}
S.~Catani and B.~R.~Webber,
%``Infrared safe but infinite: Soft gluon divergences inside the physical region,''
JHEP \textbf{10} (1997), 005
doi:10.1088/1126-6708/1997/10/005
[arXiv:hep-ph/9710333 [hep-ph]].
%184 citations counted in INSPIRE as of 01 Oct 2024

\bibitem{Aglietti:2001br}
U.~Aglietti,
%``Resummed B---\ensuremath{>} X(u) lepton neutrino decay distributions to next-to-leading order,''
Nucl. Phys. B \textbf{610} (2001), 293-315
doi:10.1016/S0550-3213(01)00316-9
[arXiv:hep-ph/0104020 [hep-ph]].

\bibitem{Aglietti:2005mb}
U.~Aglietti, G.~Ricciardi and G.~Ferrera,
%``Threshold resummed spectra in B ---\ensuremath{>} X(u) l nu decays in NLO. I,''
Phys. Rev. D \textbf{74} (2006), 034004
doi:10.1103/PhysRevD.74.034004
[arXiv:hep-ph/0507285 [hep-ph]].

%\cite{Bhattacharya:2022dtm}
\bibitem{Bhattacharya:2022dtm}
A.~Bhattacharya, M.~D.~Schwartz and X.~Zhang,
%``Sudakov shoulder resummation for thrust and heavy jet mass,''
Phys. Rev. D \textbf{106} (2022) no.7, 074011
doi:10.1103/PhysRevD.106.074011
[arXiv:2205.05702 [hep-ph]].
%10 citations counted in INSPIRE as of 20 Dec 2024

%\cite{Korchemsky:1999kt}
\bibitem{Korchemsky:1999kt}
G.~P.~Korchemsky and G.~F.~Sterman,
%``Power corrections to event shapes and factorization,''
Nucl. Phys. B \textbf{555} (1999), 335-351
doi:10.1016/S0550-3213(99)00308-9
[arXiv:hep-ph/9902341 [hep-ph]].
%311 citations counted in INSPIRE as of 29 Jan 2025

%\cite{SLD:1994idb}
\bibitem{SLD:1994idb}
K.~Abe \textit{et al.} [SLD],
%``Measurement of alpha-s (M(Z)**2) from hadronic event observables at the Z0 resonance,''
Phys. Rev. D \textbf{51} (1995), 962-984
doi:10.1103/PhysRevD.51.962
[arXiv:hep-ex/9501003 [hep-ex]].
%245 citations counted in INSPIRE as of 29 Jan 2025

%\cite{Wicke:1999zz}
\bibitem{Wicke:1999zz}
D.~Wicke,
%``Energieabh\"angigkeit von Ereignisformobservablen und der starken Kopplung,''
WUB-DIS-1999-05.
%0 citations counted in INSPIRE as of 16 Jan 2025

%\cite{DELPHI:2000uri}
\bibitem{DELPHI:2000uri}
P.~Abreu \textit{et al.} [DELPHI],
%``Consistent measurements of alpha(s) from precise oriented event shape distributions,''
Eur. Phys. J. C \textbf{14} (2000), 557-584
doi:10.1007/s100520000354
[arXiv:hep-ex/0002026 [hep-ex]].
%88 citations counted in INSPIRE as of 20 Dec 2024

%\cite{ALEPH:2003obs}
\bibitem{ALEPH:2003obs}
A.~Heister \textit{et al.} [ALEPH],
%``Studies of QCD at e+ e- centre-of-mass energies between 91-GeV and 209-GeV,''
Eur. Phys. J. C \textbf{35} (2004), 457-486
doi:10.1140/epjc/s2004-01891-4
%273 citations counted in INSPIRE as of 13 Jan 2025

%\cite{OPAL:2004wof}
\bibitem{OPAL:2004wof}
G.~Abbiendi \textit{et al.} [OPAL],
%``Measurement of event shape distributions and moments in e+ e- ---\ensuremath{>} hadrons at 91-GeV - 209-GeV and a determination of alpha(s),''
Eur. Phys. J. C \textbf{40} (2005), 287-316
doi:10.1140/epjc/s2005-02120-6
[arXiv:hep-ex/0503051 [hep-ex]].
%181 citations counted in INSPIRE as of 15 Jan 2025

%\cite{L3:2004cdh}
\bibitem{L3:2004cdh}
P.~Achard \textit{et al.} [L3],
%``Studies of hadronic event structure in $e^{+} e^{-}$ annihilation from 30-GeV to 209-GeV with the L3 detector,''
Phys. Rept. \textbf{399} (2004), 71-174
doi:10.1016/j.physrep.2004.07.002
[arXiv:hep-ex/0406049 [hep-ex]].
%193 citations counted in INSPIRE as of 20 Dec 2024

%\cite{Tarasov:1980au}
\bibitem{Tarasov:1980au}
O.~V.~Tarasov, A.~A.~Vladimirov and A.~Y.~Zharkov,
%``The Gell-Mann-Low Function of QCD in the Three Loop Approximation,''
Phys. Lett. B \textbf{93} (1980), 429-432
doi:10.1016/0370-2693(80)90358-5
%1021 citations counted in INSPIRE as of 31 Dec 2024

%\cite{Larin:1993tp}
\bibitem{Larin:1993tp}
S.~A.~Larin and J.~A.~M.~Vermaseren,
%``The Three loop QCD Beta function and anomalous dimensions,''
Phys. Lett. B \textbf{303} (1993), 334-336
doi:10.1016/0370-2693(93)91441-O
[arXiv:hep-ph/9302208 [hep-ph]].
%698 citations counted in INSPIRE as of 31 Dec 2024

%\cite{vanRitbergen:1997va}
\bibitem{vanRitbergen:1997va}
T.~van Ritbergen, J.~A.~M.~Vermaseren and S.~A.~Larin,
%``The Four loop beta function in quantum chromodynamics,''
Phys. Lett. B \textbf{400} (1997), 379-384
doi:10.1016/S0370-2693(97)00370-5
[arXiv:hep-ph/9701390 [hep-ph]].
%1320 citations counted in INSPIRE as of 21 Jan 2025

%\cite{Czakon:2004bu}
\bibitem{Czakon:2004bu}
M.~Czakon,
%``The Four-loop QCD beta-function and anomalous dimensions,''
Nucl. Phys. B \textbf{710} (2005), 485-498
doi:10.1016/j.nuclphysb.2005.01.012
[arXiv:hep-ph/0411261 [hep-ph]].
%593 citations counted in INSPIRE as of 28 Jan 2025

%\cite{Herzog:2017ohr}
\bibitem{Herzog:2017ohr}
F.~Herzog, B.~Ruijl, T.~Ueda, J.~A.~M.~Vermaseren and A.~Vogt,
%``The five-loop beta function of Yang-Mills theory with fermions,''
JHEP \textbf{02} (2017), 090
doi:10.1007/JHEP02(2017)090
[arXiv:1701.01404 [hep-ph]].
%413 citations counted in INSPIRE as of 28 Jan 2025

%\cite{Bruser:2018rad}
\bibitem{Bruser:2018rad}
R.~Br\"user, Z.~L.~Liu and M.~Stahlhofen,
%``Three-Loop Quark Jet Function,''
Phys. Rev. Lett. \textbf{121} (2018) no.7, 072003
doi:10.1103/PhysRevLett.121.072003
[arXiv:1804.09722 [hep-ph]].
%70 citations counted in INSPIRE as of 23 Jan 2025

%\cite{Kelley:2011ng}
\bibitem{Kelley:2011ng}
R.~Kelley, M.~D.~Schwartz, R.~M.~Schabinger and H.~X.~Zhu,
%``The two-loop hemisphere soft function,''
Phys. Rev. D \textbf{84} (2011), 045022
doi:10.1103/PhysRevD.84.045022
[arXiv:1105.3676 [hep-ph]].
%156 citations counted in INSPIRE as of 20 Dec 2024

%\cite{Chen:2020dpk}
\bibitem{Chen:2020dpk}
W.~Chen, F.~Feng, Y.~Jia and X.~Liu,
%``Double-real-virtual and double-virtual-real corrections to the three-loop thrust soft function,''
JHEP \textbf{12} (2022), 094
doi:10.1007/JHEP12(2022)094
[arXiv:2206.12323 [hep-ph]].
%17 citations counted in INSPIRE as of 23 Jan 2025

%\cite{Baranowski:2021gxe}
\bibitem{Baranowski:2021gxe}
D.~Baranowski, M.~Delto, K.~Melnikov and C.~Y.~Wang,
%``On phase-space integrals with Heaviside functions,''
JHEP \textbf{02} (2022), 081
doi:10.1007/JHEP02(2022)081
[arXiv:2111.13594 [hep-ph]].
%18 citations counted in INSPIRE as of 05 Jan 2025

%\cite{Baranowski:2022khd}
\bibitem{Baranowski:2022khd}
D.~Baranowski, M.~Delto, K.~Melnikov and C.~Y.~Wang,
%``Same-hemisphere three-gluon-emission contribution to the zero-jettiness soft function at N3LO QCD,''
Phys. Rev. D \textbf{106} (2022) no.1, 014004
doi:10.1103/PhysRevD.106.014004
[arXiv:2204.09459 [hep-ph]].
%19 citations counted in INSPIRE as of 23 Jan 2025

%\cite{Moult:2022xzt}
\bibitem{Moult:2022xzt}
I.~Moult, H.~X.~Zhu and Y.~J.~Zhu,
%``The four loop QCD rapidity anomalous dimension,''
JHEP \textbf{08} (2022), 280
doi:10.1007/JHEP08(2022)280
[arXiv:2205.02249 [hep-ph]].
%54 citations counted in INSPIRE as of 24 Jan 2025

%\cite{Duhr:2022yyp}
\bibitem{Duhr:2022yyp}
C.~Duhr, B.~Mistlberger and G.~Vita,
%``Four-Loop Rapidity Anomalous Dimension and Event Shapes to Fourth Logarithmic Order,''
Phys. Rev. Lett. \textbf{129} (2022) no.16, 162001
doi:10.1103/PhysRevLett.129.162001
[arXiv:2205.02242 [hep-ph]].
%66 citations counted in INSPIRE as of 24 Jan 2025

%\cite{Duhr:2022cob}
\bibitem{Duhr:2022cob}
C.~Duhr, B.~Mistlberger and G.~Vita,
%``Soft integrals and soft anomalous dimensions at N$^{3}$LO and beyond,''
JHEP \textbf{09} (2022), 155
doi:10.1007/JHEP09(2022)155
[arXiv:2205.04493 [hep-ph]].
%33 citations counted in INSPIRE as of 24 Jan 2025

%\cite{Baranowski:2024ysi}
\bibitem{Baranowski:2024ysi}
D.~Baranowski, M.~Delto, K.~Melnikov, A.~Pikelner and C.~Y.~Wang,
%``Triple real-emission contribution to the zero-jettiness soft function at N3LO in QCD,''
[arXiv:2412.14001 [hep-ph]].
%1 citations counted in INSPIRE as of 23 Jan 2025

%\cite{Baranowski:2024vxg}
\bibitem{Baranowski:2024vxg}
D.~Baranowski, M.~Delto, K.~Melnikov, A.~Pikelner and C.~Y.~Wang,
%``Zero-jettiness soft function to third order in perturbative QCD,''
[arXiv:2409.11042 [hep-ph]].
%4 citations counted in INSPIRE as of 23 Jan 2025

%\cite{Moch:2004pa}
\bibitem{Moch:2004pa}
S.~Moch, J.~A.~M.~Vermaseren and A.~Vogt,
%``The Three loop splitting functions in QCD: The Nonsinglet case,''
Nucl. Phys. B \textbf{688} (2004), 101-134
doi:10.1016/j.nuclphysb.2004.03.030
[arXiv:hep-ph/0403192 [hep-ph]].
%1403 citations counted in INSPIRE as of 20 Jan 2025

%\cite{Henn:2019swt}
\bibitem{Henn:2019swt}
J.~M.~Henn, G.~P.~Korchemsky and B.~Mistlberger,
%``The full four-loop cusp anomalous dimension in $\mathcal{N}=4$ super Yang-Mills and QCD,''
JHEP \textbf{04} (2020), 018
doi:10.1007/JHEP04(2020)018
[arXiv:1911.10174 [hep-th]].
%207 citations counted in INSPIRE as of 14 Jan 2025

%\cite{vonManteuffel:2020vjv}
\bibitem{vonManteuffel:2020vjv}
A.~von Manteuffel, E.~Panzer and R.~M.~Schabinger,
%``Cusp and collinear anomalous dimensions in four-loop QCD from form factors,''
Phys. Rev. Lett. \textbf{124} (2020) no.16, 162001
doi:10.1103/PhysRevLett.124.162001
[arXiv:2002.04617 [hep-ph]].
%169 citations counted in INSPIRE as of 03 Jan 2025

%\cite{Herzog:2018kwj}
\bibitem{Herzog:2018kwj}
F.~Herzog, S.~Moch, B.~Ruijl, T.~Ueda, J.~A.~M.~Vermaseren and A.~Vogt,
%``Five-loop contributions to low-N non-singlet anomalous dimensions in QCD,''
Phys. Lett. B \textbf{790} (2019), 436-443
doi:10.1016/j.physletb.2019.01.060
[arXiv:1812.11818 [hep-ph]].
%74 citations counted in INSPIRE as of 31 Dec 2024

%\cite{Baikov:2009bg}
\bibitem{Baikov:2009bg}
P.~A.~Baikov, K.~G.~Chetyrkin, A.~V.~Smirnov, V.~A.~Smirnov and M.~Steinhauser,
%``Quark and gluon form factors to three loops,''
Phys. Rev. Lett. \textbf{102} (2009), 212002
doi:10.1103/PhysRevLett.102.212002
[arXiv:0902.3519 [hep-ph]].
%276 citations counted in INSPIRE as of 27 Jan 2025

%\cite{Lee:2010cga}
\bibitem{Lee:2010cga}
R.~N.~Lee, A.~V.~Smirnov and V.~A.~Smirnov,
%``Analytic Results for Massless Three-Loop Form Factors,''
JHEP \textbf{04} (2010), 020
doi:10.1007/JHEP04(2010)020
[arXiv:1001.2887 [hep-ph]].
%165 citations counted in INSPIRE as of 11 Dec 2024

%\cite{Gehrmann:2010ue}
\bibitem{Gehrmann:2010ue}
T.~Gehrmann, E.~W.~N.~Glover, T.~Huber, N.~Ikizlerli and C.~Studerus,
%``Calculation of the quark and gluon form factors to three loops in QCD,''
JHEP \textbf{06} (2010), 094
doi:10.1007/JHEP06(2010)094
[arXiv:1004.3653 [hep-ph]].
%325 citations counted in INSPIRE as of 27 Jan 2025

%\cite{Lee:2022nhh}
\bibitem{Lee:2022nhh}
R.~N.~Lee, A.~von Manteuffel, R.~M.~Schabinger, A.~V.~Smirnov, V.~A.~Smirnov and M.~Steinhauser,
%``Quark and Gluon Form Factors in Four-Loop QCD,''
Phys. Rev. Lett. \textbf{128} (2022) no.21, 212002
doi:10.1103/PhysRevLett.128.212002
[arXiv:2202.04660 [hep-ph]].
%55 citations counted in INSPIRE as of 22 Jan 2025

%\cite{Ju:2021lah}
\bibitem{Ju:2021lah}
W.~L.~Ju and M.~Sch\"onherr,
%``The q$_{T}$ and \ensuremath{\Delta}\ensuremath{\phi} spectra in W and Z production at the LHC at N$^{3}$LL'+N$^{2}$LO,''
JHEP \textbf{10} (2021), 088
doi:10.1007/JHEP10(2021)088
[arXiv:2106.11260 [hep-ph]].
%45 citations counted in INSPIRE as of 08 Jan 2025

\end{thebibliography}
\end{document}